\documentclass[12pt]{article}
\usepackage{hyperref}
\usepackage{graphicx}
\usepackage{amsmath}
\usepackage{amsthm}
\usepackage{amssymb}
\usepackage{tipa}
\usepackage{subfig} 
\usepackage{xcolor}
\usepackage{mathrsfs}
\theoremstyle{plain}

\usepackage{tikz}
\usepackage{tikz-cd}
\usetikzlibrary{arrows}
\usetikzlibrary{intersections}
\usetikzlibrary{shapes.geometric}
\usetikzlibrary{decorations.pathreplacing}
\usetikzlibrary{decorations.pathmorphing, patterns,shapes}

\usepackage[nosort]{cite}

\newlength{\abstractwidth}
\renewcommand{\title}[1]{\vbox{\center\bf{\Large{#1}}}\vspace{5mm}}
\renewcommand{\author}[1]{\vbox{\center#1}\vspace{5mm}}
\newcommand{\address}[1]{\vbox{\center\em#1}}
\newcommand{\email}[1]{\vbox{\center\tt#1}\vspace{5mm}}
  
\renewcommand{\bar}{\overline}
\renewcommand{\tilde}{\widetilde}

\renewcommand{\leq}{\leqslant}
\renewcommand{\geq}{\geqslant}
\renewcommand{\Re}{\operatorname{Re}}
\renewcommand{\Im}{\operatorname{Im}}

\newcommand{\Tr}{\operatorname{tr}}
\newcommand{\nn}{\nonumber}
\newcommand{\Pf}{\operatorname{Pf}}
\newcommand{\const}{\operatorname{const}}

\newcommand{\SL}{\operatorname{SL}}

\newcommand{\CC}{\mathbb{C}}
\newcommand{\RR}{\mathbb{R}}

\newcommand{\ZZ}{\mathbb{Z}}

\newcommand{\calD}{\mathcal{D}}

\newcommand{\calO}{\mathcal{O}}

\newcommand{\TFD}{\mathrm{TFD}}
\newcommand{\Diff}{\operatorname{Diff}}

\newcommand{\Sch}{\operatorname{Sch}}

\begin{document}

\begin{titlepage}
\begin{center}
\hfill \\
\hfill \\
\vskip 1cm

\title{Spread of entanglement in a Sachdev-Ye-Kitaev chain}

\author{Yingfei Gu, Andrew Lucas and Xiao-Liang Qi}
\address{
Department of Physics, Stanford University, Stanford, CA 94305, USA
}
\email{yfgu@stanford.edu, ajlucas@stanford.edu, xlqi@stanford.edu}
\end{center}

\begin{abstract}
We study the spread of R\'enyi entropy between two halves of a Sachdev-Ye-Kitaev (SYK) chain of Majorana fermions,  prepared in a thermofield double (TFD) state.  The SYK chain model is a model of chaotic many-body systems, which describes a one-dimensional lattice of Majorana fermions, with spatially local random quartic interaction. We find that for integer R\'enyi index $n>1$, the R\'enyi entanglement entropy saturates at a parametrically smaller value than expected.   This implies that the TFD state of the SYK chain does not rapidly thermalize, despite being maximally chaotic:  instead, it rapidly approaches a prethermal state.   We compare our results to the signatures of thermalization observed in other quenches in the SYK model, and to intuition from nearly-$\mathrm{AdS}_2$ gravity.

\vspace{1in}

\today 
\end{abstract}

\end{titlepage}

\tableofcontents

\section{Introduction}

Intuition from statistical mechanics suggests that generic interacting sytsems should thermalize.  For an isolated quantum system, this seems counter-intuitive since a pure state always stays pure under unitary time evolution. Nevertheless, there is a deep sense in which a highly excited pure state can nevertheless ``look thermal".  If we study the reduced density matrix of a small subregion of the total quantum system, we expect that in a generic thermalizing quantum system, the reduced density matrix of such a region is very close to a thermal density matrix \cite{deutsch1991quantum, srednicki1994chaos, rigol2007thermalization}.  Such a thermal reduced density matrix has entanglement entropy proportional to the volume of the subregion, in contrast to the vanishing entanglement {entropy} of a pure state. Therefore thermalization of an isolated system is fundamentally related to the dynamics of the entanglement entropy between subsystems.

A direct probe of thermalization is thus to start with a 
{highly excited state with low entanglement}, and to evolve it forward in time.   For example, suppose that we start in the ground state $|\Psi\rangle$ of a quantum system for $t<0$, and at $t=0$ abruptly change the Hamiltonian so that $|\Psi\rangle$ is now highly excited.   By studying the entanglement growth in a subregion of the quantum system after such a quench, we learn how the system thermalizes;  as we have seen, the spread of entanglement is necessary for thermalization.  In many strongly interacting quantum systems, it is observed that (for large enough regions) the rate of change of the (von Neumann) entanglement entropy of a region of surface area $A$ is given by \begin{equation}
\frac{dS_{\mathrm{E}}}{dt} = s_{\mathrm{th}} \times v_{\mathrm{E}}A,  \label{eq:dSEdt}
\end{equation}
where $s_{\mathrm{th}}$ is the entropy density of the resulting thermal state, and $v_{\mathrm{E}}$ is an ``entanglement velocity" \cite{hartman2013time,
liu2014entanglement,
liu2014entanglement2}.  The entanglement velocity gives a simple measure of how rapidly the density matrix appears thermal, and -- at least by dimensional analysis -- defines a velocity scale for the thermalization of an interacting quantum system.

Another perspective on thermalization arises from quantum chaos.   In a chaotic system, quantum information present in a small subregion at time $t=0$ becomes spread out quickly.   Consider an operator $\mathcal{O}_x$ with support near a point $x$ at $t=0$.   After time evolution, the operator $\mathcal{O}_x(t) = {e^{iHt} \mathcal{O}_x e^{-iHt}}$ can become a `large' operator with support in a ball of radius $\propto t$ \cite{liebrobinson}.  The process by which these operators become delocalized, and so one must look at a large region of the system to recover a small amount of information, is coined  `scrambling' \cite{sekino}.   The spatial dynamics of scrambling is governed by a different velocity scale, called the butterfly velocity $v_{\mathrm{B}}$ \cite{shenker2014black}.   In order for a large subregion of a quantum system to thermalize, certainly operators localized inside of the subregion at $t=0$ must begin to extend outside of the subregion by the thermalization time.

Thus, the dynamics of entanglement cannot be entirely independent from the dynamics of scrambling:  both are intimately connected with thermalization.  There are plausible arguments \cite{lrbutterfly, mezei1} that scrambling should (begin) to occur \emph{first}:  \begin{equation}
v_{\mathrm{E}} \leq v_{\mathrm{B}}.  \label{eq:vEvBineq}
\end{equation}
because the growth of entanglement is impossible without the spread of information.\footnote{One can prove \cite{hartmanspeed} that $v_{\mathrm{E}} \leq v_{\mathrm{LR}}$, the Lieb-Robinson velocity \cite{liebrobinson}, though this is in general a far weaker bound \cite{lrbutterfly}.}  However, many open questions remain.  Is (\ref{eq:dSEdt}) a universal property of chaotic, thermalizing quantum systems?   If so, is $v_{\mathrm{E}}$ a physical speed in the quantum system, and if it is, does something locally well-defined propagate at this speed?  What velocity scale -- if either -- limits the onset of classical hydrodynamics, and could thus bound sound speeds and diffusion constants \cite{hartnoll2017bound}, and why?  
While many of these questions become quite delicate for many-body localized systems \cite{nandkishore} (or systems close to a many-body localization transition) \cite{xie2017, fan2017out, chenlogarithm, debanjan},  they are also not well understood for highly chaotic systems.

In this paper, we will focus on the generalized Sachdev-Ye-Kitaev (SYK) models \cite{sachdevye,kitaev2015simple, stanford1604,gu2016local} as a solvable model of a chaotic system.   The solvability of this model, and various generalizations of it \cite{gu2016local, sachdev2015bekenstein, 
davison2017thermoelectric,
gross2017generalization,
banerjee2017solvable,
bi2017instability,
gu2017chain,
song2017strongly,
chen2017tunable,
chen2017competition,
jian2017solvable,
fu2017supersymmetric,
berkooz2017higher,
turiaci2017towards,
witten2016syk,
klebanov2017uncolored,
murugan2017more
}, allows for detailed studies of quantum chaos and thermalization. Here we will initiate a study of the spatial dynamics of entanglement in the SYK chain model proposed in Ref. \cite{gu2016local}.We will describe specific details of this model later.  For now, let us simply emphasize that it is a highly disordered quantum system with a single energy scale $J$, and a large number $N$ of degrees of freedom per lattice site.   At temperatures $T\ll J$ (commonly denoted $\beta J \gg 1$, with inverse temperature $\beta=1/T$), the SYK model exhibits near-conformal invariance, becomes maximally chaotic in its early time behavior, as measured by the Lyapunov exponent, and has exponentially many low-lying excited states, leading to a non-zero entropy at zero temperature if we first take the large $N$ limit.  Due to being maximally chaotic, we would anticipate that the SYK chain is an effective thermalizer.

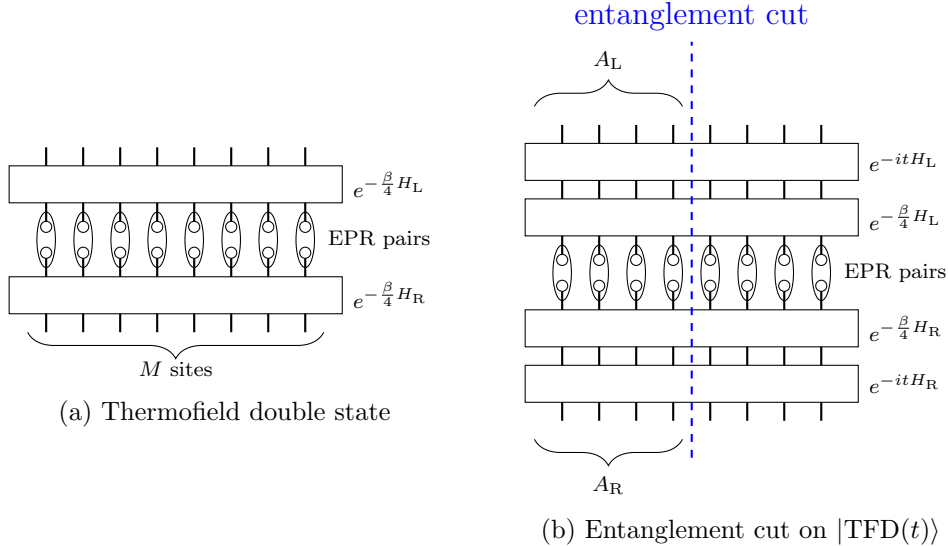
\begin{figure}
[t]
\center
\subfloat[Thermofield double state]{
\begin{tikzpicture}[scale=0.7, baseline={(current bounding box.center)}]
\draw[white] (-100pt,-70pt) rectangle (100pt,70pt);
\filldraw[fill=white] (0pt,7pt) circle (3pt);
\filldraw[fill=white] (0pt,-7pt) circle (3pt);
\draw (0pt,0pt) ellipse (5pt and 15pt) ;

\filldraw[fill=white] (20pt,7pt) circle (3pt);
\filldraw[fill=white] (20pt,-7pt) circle (3pt);
\draw (20pt,0pt) ellipse (5pt and 15pt) ;

\filldraw[fill=white] (40pt,7pt) circle (3pt);
\filldraw[fill=white] (40pt,-7pt) circle (3pt);
\draw (40pt,0pt) ellipse (5pt and 15pt) ;

\filldraw[fill=white] (60pt,7pt) circle (3pt);
\filldraw[fill=white] (60pt,-7pt) circle (3pt);
\draw (60pt,0pt) ellipse (5pt and 15pt) ;

\filldraw[fill=white] (-20pt,7pt) circle (3pt);
\filldraw[fill=white] (-20pt,-7pt) circle (3pt);
\draw (-20pt,0pt) ellipse (5pt and 15pt) ;

\filldraw[fill=white] (-40pt,7pt) circle (3pt);
\filldraw[fill=white] (-40pt,-7pt) circle (3pt);
\draw (-40pt,0pt) ellipse (5pt and 15pt) ;

\filldraw[fill=white] (-60pt,7pt) circle (3pt);
\filldraw[fill=white] (-60pt,-7pt) circle (3pt);
\draw (-60pt,0pt) ellipse (5pt and 15pt) ;

\filldraw[fill=white] (-80pt,7pt) circle (3pt);
\filldraw[fill=white] (-80pt,-7pt) circle (3pt);
\draw (-80pt,0pt) ellipse (5pt and 15pt) ;

\draw[thick] (-80pt,10pt)--(-80pt,20pt); 
\draw[thick] (-60pt,10pt)--(-60pt,20pt); 
\draw[thick] (-40pt,10pt)--(-40pt,20pt); 
\draw[thick] (-20pt,10pt)--(-20pt,20pt); 
\draw[thick] (0pt,10pt)--(0pt,20pt); 
\draw[thick] (20pt,10pt)--(20pt,20pt); 
\draw[thick] (40pt,10pt)--(40pt,20pt); 
\draw[thick] (60pt,10pt)--(60pt,20pt); 

\draw[thick] (-80pt,-10pt)--(-80pt,-20pt); 
\draw[thick] (-60pt,-10pt)--(-60pt,-20pt); 
\draw[thick] (-40pt,-10pt)--(-40pt,-20pt); 
\draw[thick] (-20pt,-10pt)--(-20pt,-20pt); 
\draw[thick] (0pt,-10pt)--(0pt,-20pt); 
\draw[thick] (20pt,-10pt)--(20pt,-20pt); 
\draw[thick] (40pt,-10pt)--(40pt,-20pt); 
\draw[thick] (60pt,-10pt)--(60pt,-20pt);

\draw (-100pt,20pt) rectangle (80pt,40pt);
\draw (-100pt,-20pt) rectangle (80pt,-40pt);
\node[right] at (80pt,30pt) {\scriptsize $e^{-\frac{\beta}{4} H_{\mathrm L}}$};
\node[right] at (80pt,-30pt) {\scriptsize $e^{-\frac{\beta}{4} H_{\mathrm R}}$};

\draw[thick] (-80pt,50pt)--(-80pt,40pt); 
\draw[thick] (-60pt,50pt)--(-60pt,40pt); 
\draw[thick] (-40pt,50pt)--(-40pt,40pt); 
\draw[thick] (-20pt,50pt)--(-20pt,40pt); 
\draw[thick] (0pt,50pt)--(0pt,40pt); 
\draw[thick] (20pt,50pt)--(20pt,40pt); 
\draw[thick] (40pt,50pt)--(40pt,40pt); 
\draw[thick] (60pt,50pt)--(60pt,40pt); 

\draw[thick] (-80pt,-50pt)--(-80pt,-40pt); 
\draw[thick] (-60pt,-50pt)--(-60pt,-40pt); 
\draw[thick] (-40pt,-50pt)--(-40pt,-40pt); 
\draw[thick] (-20pt,-50pt)--(-20pt,-40pt); 
\draw[thick] (0pt,-50pt)--(0pt,-40pt); 
\draw[thick] (20pt,-50pt)--(20pt,-40pt); 
\draw[thick] (40pt,-50pt)--(40pt,-40pt); 
\draw[thick] (60pt,-50pt)--(60pt,-40pt); 
\draw [decorate, decoration={brace,amplitude=10pt}]
(70pt,-50pt) -- (-90pt,-50pt) ;
\node at (-10pt,-70pt) {\scriptsize $M$ sites};
\node at (100pt,0pt) {\scriptsize EPR pairs};
\end{tikzpicture}
}
\hspace{20pt}
\subfloat[Entanglement cut on $|\TFD(t) \rangle$]{
\begin{tikzpicture}[scale=0.7, baseline={(current bounding box.center)}]
\filldraw[fill=white] (0pt,7pt) circle (3pt);
\filldraw[fill=white] (0pt,-7pt) circle (3pt);
\draw (0pt,0pt) ellipse (5pt and 15pt) ;

\filldraw[fill=white] (20pt,7pt) circle (3pt);
\filldraw[fill=white] (20pt,-7pt) circle (3pt);
\draw (20pt,0pt) ellipse (5pt and 15pt) ;

\filldraw[fill=white] (40pt,7pt) circle (3pt);
\filldraw[fill=white] (40pt,-7pt) circle (3pt);
\draw (40pt,0pt) ellipse (5pt and 15pt) ;

\filldraw[fill=white] (60pt,7pt) circle (3pt);
\filldraw[fill=white] (60pt,-7pt) circle (3pt);
\draw (60pt,0pt) ellipse (5pt and 15pt) ;

\filldraw[fill=white] (-20pt,7pt) circle (3pt);
\filldraw[fill=white] (-20pt,-7pt) circle (3pt);
\draw (-20pt,0pt) ellipse (5pt and 15pt) ;

\filldraw[fill=white] (-40pt,7pt) circle (3pt);
\filldraw[fill=white] (-40pt,-7pt) circle (3pt);
\draw (-40pt,0pt) ellipse (5pt and 15pt) ;

\filldraw[fill=white] (-60pt,7pt) circle (3pt);
\filldraw[fill=white] (-60pt,-7pt) circle (3pt);
\draw (-60pt,0pt) ellipse (5pt and 15pt) ;

\filldraw[fill=white] (-80pt,7pt) circle (3pt);
\filldraw[fill=white] (-80pt,-7pt) circle (3pt);
\draw (-80pt,0pt) ellipse (5pt and 15pt) ;

\draw[thick] (-80pt,10pt)--(-80pt,20pt); 
\draw[thick] (-60pt,10pt)--(-60pt,20pt); 
\draw[thick] (-40pt,10pt)--(-40pt,20pt); 
\draw[thick] (-20pt,10pt)--(-20pt,20pt); 
\draw[thick] (0pt,10pt)--(0pt,20pt); 
\draw[thick] (20pt,10pt)--(20pt,20pt); 
\draw[thick] (40pt,10pt)--(40pt,20pt); 
\draw[thick] (60pt,10pt)--(60pt,20pt); 

\draw[thick] (-80pt,-10pt)--(-80pt,-20pt); 
\draw[thick] (-60pt,-10pt)--(-60pt,-20pt); 
\draw[thick] (-40pt,-10pt)--(-40pt,-20pt); 
\draw[thick] (-20pt,-10pt)--(-20pt,-20pt); 
\draw[thick] (0pt,-10pt)--(0pt,-20pt); 
\draw[thick] (20pt,-10pt)--(20pt,-20pt); 
\draw[thick] (40pt,-10pt)--(40pt,-20pt); 
\draw[thick] (60pt,-10pt)--(60pt,-20pt);

\draw (-100pt,20pt) rectangle (80pt,40pt);
\draw (-100pt,-20pt) rectangle (80pt,-40pt);
\node[right] at (80pt,30pt) {\scriptsize $e^{-\frac{\beta}{4} H_{\mathrm L}}$};
\node[right] at (80pt,-30pt) {\scriptsize $e^{-\frac{\beta}{4} H_{\mathrm R}}$};

\draw[thick] (-80pt,50pt)--(-80pt,40pt); 
\draw[thick] (-60pt,50pt)--(-60pt,40pt); 
\draw[thick] (-40pt,50pt)--(-40pt,40pt); 
\draw[thick] (-20pt,50pt)--(-20pt,40pt); 
\draw[thick] (0pt,50pt)--(0pt,40pt); 
\draw[thick] (20pt,50pt)--(20pt,40pt); 
\draw[thick] (40pt,50pt)--(40pt,40pt); 
\draw[thick] (60pt,50pt)--(60pt,40pt); 

\draw[thick] (-80pt,-50pt)--(-80pt,-40pt); 
\draw[thick] (-60pt,-50pt)--(-60pt,-40pt); 
\draw[thick] (-40pt,-50pt)--(-40pt,-40pt); 
\draw[thick] (-20pt,-50pt)--(-20pt,-40pt); 
\draw[thick] (0pt,-50pt)--(0pt,-40pt); 
\draw[thick] (20pt,-50pt)--(20pt,-40pt); 
\draw[thick] (40pt,-50pt)--(40pt,-40pt); 
\draw[thick] (60pt,-50pt)--(60pt,-40pt);

\draw (-100pt,50pt) rectangle (80pt,70pt);
\node[right] at (80pt,60pt) {\scriptsize $e^{-i t H_{\mathrm L}}$};
\draw (-100pt,-50pt) rectangle (80pt,-70pt); 
\node[right] at (80pt,-60pt) {\scriptsize $e^{-i t H_{\mathrm R}}$};

\node at (100pt,0pt) {\scriptsize EPR pairs};

\draw[dashed, thick, blue] (-10pt,-100pt) -- (-10pt,125pt) node[above] {entanglement cut};

\draw[thick] (-80pt,70pt)--(-80pt,80pt); 
\draw[thick] (-60pt,70pt)--(-60pt,80pt); 
\draw[thick] (-40pt,70pt)--(-40pt,80pt); 
\draw[thick] (-20pt,70pt)--(-20pt,80pt); 
\draw[thick] (0pt,70pt)--(0pt,80pt); 
\draw[thick] (20pt,70pt)--(20pt,80pt); 
\draw[thick] (40pt,70pt)--(40pt,80pt); 
\draw[thick] (60pt,70pt)--(60pt,80pt); 

\draw[thick] (-80pt,-70pt)--(-80pt,-80pt); 
\draw[thick] (-60pt,-70pt)--(-60pt,-80pt); 
\draw[thick] (-40pt,-70pt)--(-40pt,-80pt); 
\draw[thick] (-20pt,-70pt)--(-20pt,-80pt); 
\draw[thick] (0pt,-70pt)--(0pt,-80pt); 
\draw[thick] (20pt,-70pt)--(20pt,-80pt); 
\draw[thick] (40pt,-70pt)--(40pt,-80pt); 
\draw[thick] (60pt,-70pt)--(60pt,-80pt); 

\draw [decorate, decoration={brace,amplitude=10pt}]
(-15pt,-90pt) -- (-95pt,-90pt) ;
\node at (-55pt,-115pt) {\scriptsize $A_{\rm R}$};
\draw [decorate, decoration={brace,amplitude=10pt,mirror}]
(-15pt,90pt) -- (-95pt,90pt) ;
\node at (-55pt,115pt) {\scriptsize $A_{\rm L}$};
\end{tikzpicture}
}

\caption{(a) Illustration of the TFD state, which is obtained by applying imaginary time evolution $e^{-\frac{\beta}4H_L}$ and $e^{-\frac{\beta}4H_R}$ to a state $|I\rangle$ of the two-chain system. $|I\rangle$ is a direct product of local EPR pairs between the two sites in the two chains at the same spatial location.
(b) The real time evolution of the TFD state by $U(t)= \exp [-i t (H_{\mathrm L}+H_{\mathrm R}) ]$ and our choice of entanglement cut. We study the Renyi entropies of the region $A=A_L\cup A_R$, with support on both chains.}
\label{fig: TFD}
\end{figure}

More specifically, we study the entropy growth in the generalized SYK model after a global quench {from} a special initial state, the thermofield double (TFD) state \cite{israel1976thermo}. 
To construct the TFD, we tensor product two copies of the original Hilbert space of the SYK chain: $\mathcal{H} = \mathcal{H}_{\mathrm{L}}\otimes \mathcal{H}_{\mathrm{R}}$.   We have denoted the copies as left (L) and right (R).  The TFD state at time $t=0$ is \begin{equation}
|\mathrm{TFD}\rangle \propto e^{-\beta (H_{\mathrm{L}}+H_{\mathrm{R}})/4}|I\rangle.   \label{eq:TFDintro}
\end{equation}
The state $|I\rangle$ is a direct product of local EPR pairs between the two systems L and R (See. Fig. \ref{fig: TFD}). For a spatial entanglement cut, $|I\rangle$ has no entanglement entropy.
$H_{\mathrm{L}}$ and $H_{\mathrm{R}}$ are suitable notions of the SYK-chain Hamiltonian acting on only the L or R degrees of freedom (see Section \ref{sec:TFD} for precise definitions).
Upon tracing out either the L or the R degrees of freedom, the resulting density matrix is thermal.
However, we may define a Hamiltonian for the combined LR system such that $|\mathrm{TFD}\rangle$ is not an eigenstate.  If we look at a suitable subregion $A$ with support in both the L and R chains, 
such as the left half of the chain shown in Fig. \ref{fig: TFD}, we can observe the spatial spread of entanglement in this doubled system.

More specifically, by using the replica trick, we compute the $n$-th R\'enyi entanglement entropy $S_{A,n}$ (with integer $n>1$) of the reduced density matrix of $|\mathrm{TFD}(t)\rangle$ in one half of the TFD chain.  When the coupling between neighboring sites is the smallest energy scale, 
 we find that the entropy increases linearly in time as in (\ref{eq:dSEdt}), with the growth rate 
\begin{equation}
\frac{dS_{A,n}}{dt} \propto T.  \label{eq:linearvE}
\end{equation}
The linear growth slows down at long time and eventually has to saturate if the length of the chain is finite. We study the late time behavior by two different approaches. The first approach is a perturbative expansion in the coupling strength between neighboring sites, and allows us to compute the onset of deviation from linear growth at an intermediate time scale. Such an approach does not apply directly to the long time limit.  In the second approach, we make a simple ansatz for correlation functions that  allows us to compute the entropy at all times by a much simpler (but still nonlinear) geometric problem. Solving this problem in the long real time limit predicts a saturation of the R\'enyi entropy. Since we have done a restricted minimization of action, what we obtain is an upper limit of the entropy:  
\begin{equation}
\frac{S_{A,n}(t=\infty)}{\frac{1}{2} M} \leq\frac{n}{n-1} \frac{c_vT}{2},  \label{eq:introsat}
\end{equation}
Here $M$ is the length of the chain, so that $\frac{S_{A,n}(t=\infty)}{\frac{1}{2}M}$ is the entropy per site. $c_v$ is the specific heat of the doubled SYK chain, which is a constant at low temperature. Surprisingly, for $n>1$, at low temperature the R\'enyi entropy density upper limit is \emph{parametrically} smaller than that of the thermal ensemble, which implies there are degrees of freedom that do not thermalize in the large $N$ limit.   

We propose that this phenomenon of prethermalization is related to the presence of a large density of almost localized states at very low energy.  These states are responsible for the zero temperature entropy. They evolve slowly and do not contribute to entanglement if we first take the large $N$ limit. As a consequence, the entropy growth is upper bounded by $c_vT = s_{\mathrm{th}}(T)  -s_{\mathrm{th}}(0)$:  the change in entropy due to finite temperature $T$.   This does not include the non-vanishing zero temperature entropy density of the SYK chain.  We expect the system eventually thermalizes but that the thermalization time diverges in the large $N$ limit. The fact that the SYK chain only prethermalizes rapidly, but thermalizes slowly, implies that Eq. (\ref{eq:dSEdt}) may not be a sensible definition of an entanglement velocity.  We will discuss this possibility in much more detail at the end of the paper.   If one uses the definition (\ref{eq:dSEdt}) for $v_{\mathrm{E}}$, then using (\ref{eq:linearvE}) we find that $v_{\mathrm{E}} \propto T$ for the SYK chain.

It is interesting to discuss the relation of our results with holographic models of strongly interacting theories.  Although the holographic dual of the  SYK  model is not known,  the SYK model shares many similar properties with holographic models containing extremal horizons \cite{jensen, maldacena2016conformal, traversable}.  We show that these holographic models  exhibit similar `early' time von Neumann entanglement growth, with $dS_{\mathrm{E}}/dt \propto T$, but the von Neumann entanglement $S_{\mathrm{E}}$ saturates at the thermal value, including the extremal zero temperature contribution.   Although we are unable to explicitly compute the analogous holographic R\'enyi entropy growth, we propose that the R\'enyi entropy in an analogous holographic setting  may behave similarly to the SYK chain.  This arises due to subtleties with gravitational dynamics in $\mathrm{AdS}_2$.

Finally, we note that two models studying thermalization in (single site) SYK models in a somewhat different context have recently appeared \cite{eberlein2017quantum, kourkoulou2017pure}; both studies show evidence for rapid thermalization.   Evidence for eigenstate thermalization in the SYK model has also recently appeared \cite{sonner}. Our results are not inconsistent with theirs, but we defer a detailed comparison to the end of the paper.

The rest of the paper is organized as follows: in Sec.~\ref{sec: setups} we present the explicit setup of the global quench problem we are studying.  We compute $S_{A,n}(t)$ in the limit where the coupling between different sites is much weaker than the on-site coupling in Sec.~\ref{sec: weak link limit}, and in Sec.~\ref{sec: beyond 1} we consider the leading perturbative correction to this result. In Sec.~\ref{sec:geo} we present a geometric interpretation of the quench problem we are studying and compute the long time saturation of the entanglement entropy. This is a regime where both previous calculations fail. 
In Sec. \ref{sec:holography} we compare our result to intuition from holography.
In Sec.~\ref{sec: conclusion and discussion} we discuss the broader implications of our result and compare to other recent studies on quantum quenches in the SYK model \cite{kourkoulou2017pure, eberlein2017quantum}.

\section{Setup}
\label{sec: setups}

\subsection{The SYK Model}
The Sachdev-Ye-Kitaev (SYK) model \cite{sachdevye,kitaev2015simple} describes $N$ Majorana fermions with quartic random all-to-all interactions.   The Hamiltonian of this model is
\begin{align}
H= \sum_{1\leq j < k < l < m \leq N} J_{jklm} \chi_j \chi_k \chi_l \chi_m , \quad \lbrace \chi_j ,\chi_k \rbrace = \delta_{jk},  \label{eq:SYKHamiltonian}
\end{align}
where $\lbrace J_{jklm}\rbrace$ are independent, mean-zero random couplings: \begin{equation}
\bar{J_{jklm}}=0, \hspace{1in} \bar{J_{jklm}^2} = \frac{3!}{N^3} J^2.
\end{equation}
The model is solvable at large $N$, and maximally chaotic at strong coupling $N\gg \beta J \gg 1$
\cite{kitaev2015simple, stanfordbound,stanford1604}.  It provides a rare example of chaotic yet tractable many-body systems.

Recently, many generalizations of the SYK model have been proposed. In particular, \cite{gu2016local} studied a higher dimensional lattice generalization of the SYK model with spatial locality. 
For a one-dimensional chain, the Hamiltonian of the generalized SYK model is given by 
\begin{align}
H= \sum_{x=1}^M\left( 
\sum_{1\leq j<k<l<m\leq N} J_{jklm,x} 
\chi_{j,x} \chi_{k,x} \chi_{l,x} \chi_{m,x} + 
\sum_{j<k, l<m} J'_{jklm,x} 
\chi_{j,x} \chi_{k,x} \chi_{l,x+1} \chi_{m,x+1} 
\right)\label{Hchain}
\end{align}
where the couplings $\lbrace J_{jklm,x} \rbrace$ and $\lbrace  J'_{jklm,x} \rbrace$ are all  independent Gaussian random variables with mean zero, and variances \begin{equation}
\bar{J_{jklm}^2}=\frac{3!}{N^3} J^2_0, \hspace{1in} \bar{{J'_{jklm}}^2} = \frac{1}{N^3} J_1^2
\end{equation}
 It is convenient to define an effective coupling constant \begin{equation}
 J= \sqrt{J_0^2+J_1^2},
 \end{equation}
  which determines the local properties of the model. 
Similar to the original SYK model, this generalized model is solvable at large $N$ and maximally chaotic at strong coupling.  At leading order in $N$,  this model has a saddle point which is equivalent to the one-site SYK model.   At next-to-leading order in $N$, there is non-trivial spatial dynamics with dynamical critical exponent $z=\infty$, also known as local criticality.   Local criticality implies that space does not scale under renormalization group flow, and is responsible for some of the particular features of the SYK chain model.

The spatial locality of the model enables us to study thermal transport and the spatial propagation of scrambling and chaos.  The out-of-time-order correlation function (OTOC)  takes the form 
\begin{equation}
\frac1{N^2}\sum_{i,j}\langle \chi_{ix}(t)\chi_{jy}(0)\chi_{ix}(t)\chi_{jy}(0)\rangle_\beta~ \propto ~{\rm const.}+\frac1Ne^{\lambda_{\mathrm{L}}\left(t-x/v_{\mathrm{B}}\right)}
\end{equation} 
with Lyapunov exponent $\lambda_{\mathrm{L}}=2\pi T$ and butterfly velocity $v_{\mathrm{B}} = \sqrt{2\pi TD}$, with $D$ the thermal diffusion constant \cite{gu2016local}.
We note that this relation between $v_{\mathrm{B}}$ and thermal diffusion constant $D$ holds in holographic locally critical theories as well \cite{blake2016universal,blake2017diffusion}.   
Charge transport can also be studied in a modified model with charge conservation \cite{davison2017thermoelectric},  though we will focus in this paper on the simpler model above.

Before describing how to exactly solve this model in the limit $N\gg \beta J \gg 1$ in Sec.~\ref{sec:SYKpath}, we would like to first define the TFD state described in the introduction, and discuss how to compute the R\'enyi entropy in this state.

\subsection{Thermofield double state and global quench}
\label{sec:TFD}
We consider two copies of a single SYK chain, and consider a special initial state: the thermofield double (TFD) state \cite{israel1976thermo}.  As we will see, analytic computations in such a doubled state are tractable; the qualitative features of  entropy growth in the quenched time evolution of a short-range entangled initial state should not depend on details of the initial state.   
  
We first give the general definition of the thermofield double state. For a system with lattice sites labeled by $x$, we first choose a basis $\left|a,x\right\rangle,~a=1,2,...,D$ on each site. Here $D$ is the Hilbert space dimension of each site ($D=2^{N/2}$ for the SYK chain model). Then we consider the following state of the doubled system, which is a direct product of maximally entangled pairs on each site:
\begin{align}
\left|I\right\rangle=\bigotimes_x\left(D^{-1/2}\sum_{a}|a,x\rangle_{\mathrm L}\otimes |a,x\rangle_{\mathrm R}\right)
\end{align}
Here L and R (left and right) label the two copies of the system. In state $|I\rangle$, each chain is maximally entangled with the other chain, but when we consider the two sites at $x$ together, they are unentangled with the rest of the chain. Indeed, interpreting the chain labels L and R as an ``internal" label, $|I\rangle$ is a direct product state with no spatial entanglement. 

For a given Hamiltonian $H$ of the original single chain problem, we can define its transpose $H^T$ by taking the matrix transpose in the basis $\left|\left\{a_x\right\}\right\rangle\equiv \otimes_x\left|a_x,x\right\rangle$.\footnote{It should be noted that transpose is a basis-dependent operation, so that it is essential to first define the basis.} More explicitly, $H^T$ is defined as
\begin{align}
H^T\equiv\sum_{\left\{a_x\right\},\left\{b_x\right\}}\left\langle \left\{a_x\right\}\right|H\left|\left\{b_x\right\}\right\rangle \left|\left\{b_x\right\}\right\rangle \left\langle \left\{a_x\right\}\right|
\end{align}
Now define a Hamiltonian in the doubled system
\begin{align}
H_{\mathrm D}=H_{\mathrm L}+H_{\mathrm R},~\text{with~}H_{\mathrm L}=H\otimes \mathbb{I},~H_{\mathrm R}=\mathbb{I}\otimes H^T
\end{align}
such that $H$ acts on the left system and $H^T$ acts on the right system.  One can explicitly check that the state $|I\rangle$ satisfies
\begin{align}
\left(H_{\mathrm L}-H_{\mathrm R}\right)|I\rangle=0\label{TFDcondition}
\end{align}
The TFD state $|I\rangle$, introduced in (\ref{eq:TFDintro}), is then  defined as
\begin{align}
|\TFD\rangle=Z_\beta^{-1/2}e^{-\frac\beta 4(H_{\mathrm L}+H_{\mathrm R})}|I\rangle 
\end{align}
with $Z_\beta={\rm tr}\left(e^{-\beta H}\right)$ the thermal partition function of the single-chain system. 

A key property of TFD is that the reduced density matrix of the L chain alone, or the R chain alone, is thermal, with inverse temperature $\beta$. This can be directly shown by applying Eq. (\ref{TFDcondition}) to obtain $|\TFD\rangle=Z_\beta^{-1/2}e^{-\frac{\beta}2H_{\mathrm L}}|I\rangle$ and use the fact that $|I\rangle$ maximally entangles the two chains. One can view the TFD state as a purification of the thermal density matrix, in which the chain R plays the role of thermal bath of chain L. Compared to a generic purification, the TFD state has the special property that the entanglement between the two chains is spatially local. The state $|I\rangle$ (which is the $\beta\rightarrow 0$ limit of $|\TFD\rangle$) has zero entanglement entropy between different spatial regions. $|\TFD\rangle$ at finite temperature is obtained by a finite time imaginary time evolution of $|I\rangle$.  This imaginary time evolution leads to spatial entanglement.  However, any resulting entanglement entropy will satisfy an area law \cite{wolf2008area}, as long as $H_{\mathrm{L,R}}$ are local. In the one-dimensional chain case, this means the entanglement entropy of a connected region $A$ stays finite even if the size of $A$ and its compliment go to infinity. 

The definition of $\left|\TFD\right\rangle$ is not unique, since it depends on a basis choice. However, different definitions lead to $|I\rangle$ that are related by a product of local unitaries, which does not change entanglement properties of the TFD state as long as the definition of $H^T$ is conjugated by these unitaries correspondingly.  For concreteness, we give an explicit definition of $|\mathrm{TFD}\rangle$ in the SYK chain case.  Denote the Majorana fermion operators by $\chi_{j,x,\mathrm{L}}$ and $\chi_{j,x,\mathrm{R}}$, with $j=1,2,...,N$.
(We remind the reader that $N$ must be even, to have a well-defined Hilbert space at each site.)
One convenient choice of the state $|I\rangle$ can be defined by the following equations:
\begin{align}
c_{j,x,\mathrm{L}}=\frac12\left(\chi_{2j-1,x,\mathrm{L}}+i\chi_{2j,x,\mathrm{L}}\right), &~c_{j,x,\mathrm{R}}=\frac12\left(\chi_{2j,x,\mathrm{R}}-i\chi_{2j-1,x,\mathrm{R}}\right)\nonumber\\
\left(c_{j,x,\mathrm{L}}-c_{j,x,\mathrm{R}}\right)|I\rangle=0,&~\left(c_{j,x,\mathrm{L}}^\dagger +c_{j,x,\mathrm{R}}^\dagger\right)|I\rangle=0
\end{align}
with $j=1,2,...,N/2$. In the eigenbasis of $n_{j,x,\mathrm{L(R)}}=c_{j,x,\mathrm{L(R)}}^\dagger c_{j,x,\mathrm{L(R)}}$, the state is a product of $|0\rangle_{\mathrm L}|1\rangle_{\mathrm R}+|1\rangle_{\mathrm L}|0\rangle_{\mathrm R}$ on each site. This leads to the equations for the Majorana operators
\begin{align}
\left(\chi_{j,x,\mathrm{L}}+i\chi_{j,x,\mathrm{R}}\right)|I\rangle=0, ~~ ( j=1,2,\ldots,N).
\end{align}
In this choice, one can check that the SYK chain Hamiltonian (\ref{Hchain}) satisfies $H^T=H$. Therefore we can take two identical chains and the TFD state is defined as\footnote{This discussion, and the relation $H^T=H$, generalize to an SYK model with $q$-body interactions \cite{stanford1604}.}
\begin{equation}
|\TFD\rangle=Z_\beta^{-1/2}\exp\left[-\frac\beta 4\left(H\otimes\mathbb{I}+\mathbb{I}\otimes H\right)\right]|I\rangle.
\end{equation}

From the perspective of time evolution, the thermofield double state is an eigen-state of operator $H \otimes I - I \otimes H^T $ but not the eigen-state of our Hamiltonian $H_{\mathrm D}= H \otimes I + I \otimes H^T$. Therefore, we can treat $H_{\mathrm D}$ as the initial state after a global quench, and apply the corresponding time evolution operator to
 obtain a time dependent state:
\begin{equation}
| \TFD (t) \rangle = U(t) | \TFD \rangle, \quad  U(t)= \exp \left[ -  i  \left( H \otimes I + I \otimes H^T \right) t  \right]
\end{equation}
Now we can look at the sub-region $A=A_{\mathrm L} \cup A_{\mathrm R}$ which is supported on two sides as shown in Fig.~\ref{fig: tfd} (a) and consider its reduced density matrix:
\begin{equation}
\rho_A(t):= \Tr_{A^C} | \TFD (t) \rangle \langle \TFD(t) |.
\end{equation}
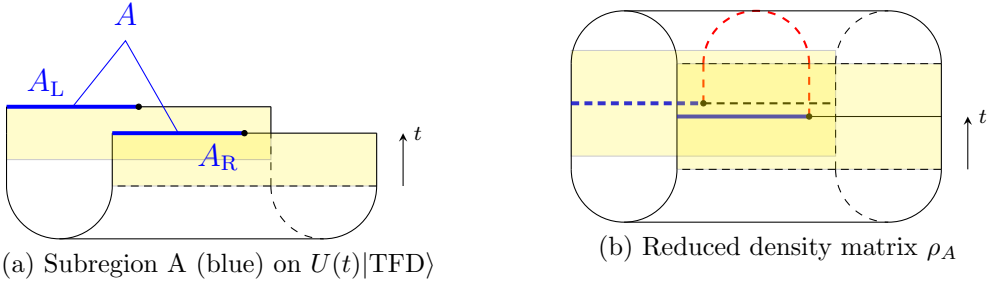
\begin{figure}[t]
\center
\subfloat[Subregion A (blue) on $U(t)| \TFD \rangle$]{
\begin{tikzpicture}[baseline={(current bounding box.center)}]

\draw (-40pt,-20pt) arc (0:-180:20pt);
\draw (-40pt,-20pt) -- (-40pt, 0pt);
\draw (-80pt,-20pt) -- (-80pt, 0pt);

\draw (60pt,-20pt) arc (0:-90:20pt);
\draw[dashed] (20pt,-20pt) arc (-180:-90:20pt);
\draw (60pt,-20pt) -- (60pt, 0pt);
\draw[dashed] (20pt,-20pt) -- (20pt, 0pt);

\draw (40pt,-40pt) -- (-60pt,-40pt);
\draw (60pt,0pt) -- (-40pt,0pt);

\filldraw[fill=yellow, opacity=0.2] (-40pt,-20pt) rectangle (60pt,0pt);
\filldraw[fill=yellow, opacity=0.2] (-80pt,10pt) rectangle (20pt,-10pt);

\draw[densely dashed] (60pt,-20pt) -- (-40pt,-20pt);
\draw [->,>=stealth] (70pt,-20pt) --(70pt,0pt) node[right]{\scriptsize $t$};

\draw[blue, line width=1.5pt] (-40pt,0pt) -- (10pt,0pt);
\filldraw[black] (10pt,0pt) circle (1pt);

\draw[blue, line width=1.5pt] (-80pt,10pt) -- (-30pt,10pt);
\draw(-30pt,10pt)--(20pt,10pt);
\draw (20pt,10pt)--(20pt,0pt);
\draw (-80pt,10pt)--(-80pt,0pt);
\filldraw[black] (-30pt,10pt) circle (1pt);

\node[blue] at (-35pt,45pt) {$A$};
\draw[blue] (-55pt,10pt) -- (-35pt,35pt)--(-15pt,0pt);
\node[blue,below] at (0pt,0pt) {$A_{\mathrm R}$};
\node[blue,above] at (-65pt,10pt) {$A_{\mathrm L}$};
\end{tikzpicture}
}
\hspace{40pt}
\subfloat[Reduced density matrix $\rho_A$]{
\begin{tikzpicture}[baseline={(current bounding box.center)}]
\draw (-40pt,20pt) arc (0:180:20pt);
\draw (-40pt,-20pt) arc (0:-180:20pt);
\draw (-40pt,-20pt) -- (-40pt, 20pt);
\draw (-80pt,-20pt) -- (-80pt, 20pt);

\draw[dashed,red,thick] (10pt,20pt) arc (0:180:20pt);
\draw[dashed,red,thick] (10pt,0pt) -- (10pt, 20pt);
\draw[dashed,red,thick] (-30pt,5pt) -- (-30pt, 20pt);

\draw (60pt,20pt) arc (0:90:20pt);
\draw (60pt,-20pt) arc (0:-90:20pt);
\draw[dashed] (20pt,20pt) arc (180:90:20pt);
\draw[dashed] (20pt,-20pt) arc (-180:-90:20pt);
\draw (60pt,-20pt) -- (60pt, 20pt);
\draw[dashed] (20pt,-20pt) -- (20pt, 20pt);

\draw (40pt,40pt) -- (-60pt,40pt);
\draw (40pt,-40pt) -- (-60pt,-40pt);

\draw (60pt,0pt) -- (-40pt,0pt);

\draw[densely dashed] (60pt,20pt) -- (-40pt,20pt);
\draw[densely dashed] (60pt,-20pt) -- (-40pt,-20pt);
\draw [->,>=stealth] (70pt,-20pt) --(70pt,0pt) node[right]{\scriptsize $t$};

\draw[blue, line width=1.5pt] (-40pt,0pt) -- (10pt,0pt);
\draw[blue, line width=1.5pt, densely dashed] (-80pt,5pt) -- (-30pt,5pt);
\draw[thick, densely dashed] (-30pt,5pt) -- (20pt,5pt);
\filldraw[black] (10pt,0pt) circle (1pt);
\filldraw[black] (-30pt,5pt) circle (1pt);

\filldraw[fill=yellow, opacity=0.2] (-80pt,25pt) rectangle (20pt,-15pt);
\filldraw[fill=yellow, opacity=0.2] (-40pt,20pt) rectangle (60pt,-20pt);
\end{tikzpicture}}
\caption{
(a) We picture the TFD state of the SYK chain model as a half tube. The two top edges correspond to the states in left and right Hilbert space. The subregion $A=A_{\mathrm L} \cup A_{\mathrm R}$ is defined on both sides, and colored blue. 
The yellow shaded region corresponds to the real time evolution $U(t)$, and the circular portion of the tube represents the initial imaginary time evolution of (\ref{eq:TFDintro}).
(b) The density matrix $\rho_A$ is then found by gluing two $|\TFD(t)\rangle$ states on $A^c$ and leaving  $A$  (blue) `open'.   This perspective is useful in computing the partition function $Z_{A,n}$, where the blue lines play the role of branch cuts.    Each replicated fermion $\chi_{\alpha}$ shifts its replica index to $\alpha+1$ when it crosses the right branch cut line (the side closer to the reader) from below and shift to $\alpha-1$ when it crosses the left branch cut line from the above.  We can further deform the two horizontal blue branch cuts to a single vertical dashed branch cut (shown in red).}
\label{fig: tfd}
\end{figure}
By construction, each chain in the TFD state has a thermal density matrix that is invariant under time evolution. Consequently a region that is a subset of only $L$ chain or only $R$ chain also has a thermal time-independent density matrix. In the region we choose, $\rho_{A_{\mathrm L}}=\Tr_{A_{\mathrm R}} \rho_A(t)$ and $\rho_{A_{\mathrm R}}$ are thermal, but $\rho_A$ is time-dependent. If region $A$ thermalizes after a long time, $\rho_A$ should approach the thermal density matrix $\rho_{A_{\mathrm L}}\otimes \rho_{A_{\mathrm R}}$.   Therefore, the  increase in entropy of region $A$ during thermalization corresponds to the decrease of mutual information $I(A_{\mathrm L}:A_{\mathrm R})=S(\rho_{A_{\mathrm L}})+S(\rho_{A_{\mathrm R}})-S(\rho_A)$ between the two regions $A_{\mathrm L}$ and $A_{\mathrm R}$. If the system thermalizes, the mutual information should vanish in long time (or at least becomes subleading in volume of $A$). Physically, the decrease of mutual information is a consequence of the scrambling of quantum information during chaotic time evolution.  Correlation between operators in $A_{\mathrm L}$ and $A_{\mathrm R}$ evolves to more and more non-local operators that cannot be revealed in region $A_{\mathrm L}$ and $A_{\mathrm R}$ \cite{hosur2016chaos,fan2017out}.

\subsection{Twist operators and R\'enyi entropy}
It is difficult to directly calculate von Neumann entropy of the region $A$. Instead we use the replica trick \cite{calabrese2004entanglement} to compute the R\'enyi entropy for the subsystem $A$ by a path integral:
\begin{align}
S_{A,n}= \frac{\log \Tr \rho_A^n}{1-n}, \quad \Tr \rho_A^n = \frac{Z_{A,n}}{Z_\beta^n},
\end{align}
where the factor of $Z_\beta^n$ in the denominator arises from the definition of the TFD state.  The numerator $Z_{A,n}$ is the partition function evaluated on a ``twisted'' manifold corresponding to an $n$-sheeted cover of spacetime. 
The easiest way to understood the partition function on this twisted manifold is to consider a partition function defined on $n$ replicas of the original theory.  For the generalized SYK model with Majorana fermions $\chi_{j,x}$, with $j=1,2,...,N$ the flavor index and $x$ the lattice site coordinate, the replicated theory describes fermions $\chi^\alpha_{j,x}$ with $\alpha=1,\ldots,n$. We then write down a path integral for all $\chi^\alpha_{j,x}$, containing evolution in Euclidean time, which prepares the TFD state, followed by evolution for real time $t$.   From the perspective of the twisted manifold, the replica index of a fermion describes which sheet of  Euclidean space-time it lives on.  These sheets are connected through the branch cut lines shown in blue in Fig. \ref{fig: tfd}, such that fermions passing across a branch cut have their replica index cyclically permuted.
More explicitly, the boundary condition of fermions is 
\begin{equation}
\chi^{\alpha}_{j,x}(\tau_*^+) =
\begin{cases}
 \chi^{\alpha+1}_{j,x}(\tau_*^-) , & x \in A_{\mathrm L}\\
 \chi^{\alpha-1}_{j,x}(\tau_*^-) , & x \in A_{\mathrm R}\\
 \chi^{\alpha}_{j,x}(\tau_*^-) , & x \in \overline{A}
 \end{cases}
\end{equation}
where $\tau_*$ is the time location of the branch cut. $Z_{A,n}$ is the partition function for replicated theory with the above boundary condition. 

The position of the branch cut line is not important; it can be moved around by relabelling fermions in different replicas. The only invariant information about the twist is the end points of the branch cut.\footnote{We can view the local permutation of different replica as a gauge transformation, and the branchcut points are gauge fluxes.}   For our convenience, we can move the branch cut points to a ``time-like" line connecting the two branch cut points (red line in Fig. \ref{fig: tfd} (b)). With this gauge choice, the boundary condition in the time direction remains untwisted, and the only effect of the twist operators is to modify the spatial couplings between fermions on different sites.  Denoting the sites separated by the boundary of $A$ as $x_*$ and $x_*+1$, the twisted coupling is between $\chi_{j,x_*}^{\alpha}(\tau)$ and $\chi_{j,x_*+1}^{\alpha+1}(\tau)$ when time $\tau$ is in the interval of the branchcut line.    The coupling is diagonal in $\alpha$ everywhere else.   

It is also helpful to write another equivalent expression of the R\'enyi entropy. If we define a twist operator $X_{An}$ which is applied to the branch cut lines $A_{\mathrm L}$ or $A_{\mathrm R}$ and cyclically permutes the replica index, the R\'enyi entropy can also be written in a time-ordered thermal two-point function of twist operators:
\begin{eqnarray}
e^{-(n-1)S_{A,n}}&=&{\rm tr}\left[ X_{An}^\dagger \left(-it\right) \left(\rho_{\beta}^{\otimes n}\right)^{1/2} X_{An}\left(it\right)\left(\rho_{\beta}^{\otimes n}\right)^{1/2}\right]\nn\\
&=&\left\langle X_{An}^\dagger\left(\frac{\beta}2-it\right)X_{An}(it)\right\rangle_\beta \label{eq: twist twopoint}
\end{eqnarray}
Here $\langle ...\rangle_\beta={\rm tr}\left(\rho_\beta^{\otimes n}...\right)$ is the thermal average in $n$ copies of the single chain system. The real time evolution and imaginary time evolution can be drawn as a contour in the complex plane of time, as shown in Fig. \ref{fig: time contour}(a). 

\subsection{The path integral for the SYK chain}\label{sec:SYKpath}
The discussion of the previous subsection was completely general.  It applies to the TFD state of any system with a spatially local Hamiltonian.  We will now write down the coupling more explicitly for the SYK chain model.

\begin{figure}[t]
\center
\begin{tikzpicture}[scale=1.0, baseline={(current bounding box.center)}]
\draw[red] (40pt,2pt) arc (0:180:40pt);
\draw (40pt,-2pt) arc (0:-180:40pt);
\draw (-40pt,-2pt) -- (-80pt, -2pt);
\draw[red] (-40pt,2pt) -- (-80pt, 2pt);
\filldraw (-80pt,0pt) circle (2pt) node[below] { $-it+\frac{\beta}{2}$};
\draw (40pt,-2pt) -- (20pt, -2pt);
\draw[red] (40pt,2pt) -- (20pt, 2pt);
\filldraw (20pt,0pt) circle (2pt) node[below] { $ it$};
\node at (40pt, 50pt) {contour $C_1$};
\node at (40pt, -50pt) {contour $C_2$};
\draw [->,>=stealth] (-100pt,0pt) -- (100pt,0pt);
\draw [->,>=stealth] (0pt,-80pt) -- (0pt,80pt);
\end{tikzpicture}
\caption{The time contour involved in the entanglement entropy calculation. This figure is plotted in complex plane $z=\exp (i\frac{2\pi}{\beta} t_{\CC})$, where $t_{\CC}=\tau+it$ is the complex time variable. The red part represents the $\langle \TFD(t)|$ in Fig.~\ref{fig: tfd}, and the black represents the $| \TFD(t) \rangle$. For later convenience, we name them as $C_1$ and $C_2$.}
\label{fig: time contour}
\end{figure}
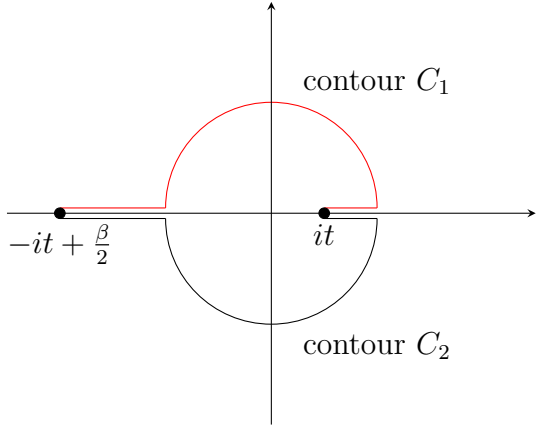

The replicated partition function $Z_{A,n}$ with the boundary condition described in the previous subsection can be written as
\begin{align}
Z_{A,n}[J]&= \int \prod_{\alpha,j, x}  \calD  \chi_{j,x}^\alpha \exp \left( - S_J[\chi_{j,x}^\alpha] \right), \nn \\ 
S_{J}[\chi_{x,j}^\alpha]&=\sum_{\alpha,x} \left[ \int_C  d\tau \sum_j \frac{1}{2} \chi_{j,x}^\alpha \partial_\tau \chi_{j,x}^\alpha  \right. \nn\\
&- \left.
 \sum_{jklm} \left( J_{jklm,x} 
\chi_{j,x}^\alpha \chi_{k,x}^\alpha \chi_{l,x}^\alpha \chi_{m,x}^\alpha +
 {J'}_{jklm,x}g_x^{\alpha\beta}(\tau)
\chi_{j,x}^\alpha \chi_{k,x}^\alpha \chi_{l,x+1}^\beta \chi_{m,x+1}^\beta \right) \right]
\end{align}
where $C$ stands for a special time contour for the thermofield double states as shown in Fig.~\ref{fig: time contour}. At the boundary of $A$, between sites $x_*$ and $x_*+1$, the contour is split into two parts $C_1$ and $C_2$ by the twist operators. The branchcut line runs along $C_1$.  The effect of the twist is to modify the spatial coupling ${J'}_{jklm,x}$ term by the matrix $g_x^{\alpha\beta}(\tau)$, given by
\begin{equation}
g_x^{\alpha\beta}(\tau)=\begin{cases}\delta^{\beta,\alpha+1},&\text{if~}\tau\in C_1~\text{and}~x=x_*\\
\delta^{\alpha\beta},& \text{elsewhere}\end{cases}
\end{equation}

In a generic system, to compute the quenched average of the R\'enyi entropy $S_{A,n}$ over disordered couplings $J_{jklm,x}$ and $J'_{jklm,x}$ one should compute $\overline{Z_{A,n}^k}$ for a general integer $k$, and then analytically continuate to the $k\rightarrow 0$ limit. In the SYK model, it is known that at leading and next-to-leading order in $N$, the partition function is replica diagonal, such that $\overline{Z_{A,n}^k}\simeq \overline{Z_{A,n}}^k$\cite{stanford1604,gu2016local,cotler2017black}. Therefore we will directly work with $\overline{Z_{A,n}}$.  The average over the Gaussian-distributed random couplings $J$ and $J'$ leads to the following partition function:
\begin{align}
\overline{Z_{A,n}}&= \int \prod_{\alpha,j, x}  \calD  \chi_{j,x}^\alpha \exp \left( - S[\chi_{j,x}^\alpha] \right), \nn \\ 
S[\chi_{x,j}^\alpha]&=\sum_{x} \left[ \int_C  d\tau \sum_{j,\alpha} \frac{1}{2} \chi_{j,x}^\alpha \partial_\tau \chi_{j,x}^\alpha  - \frac{1}{8N^3}  \iint_C d\tau_1 d\tau_2
  \left(\sum_{\alpha,\beta} J_0^2 \left(\sum_j 
\chi_{j,x}^\alpha (\tau_1) \chi_{j,x}^\beta (\tau_2) \right)^4\right. \right. \nn\\
&+ \left. \left.	
\sum_{\alpha,\beta,\alpha',\beta'} J_1^2 \left( \sum_j
\chi_{j,x}^\alpha (\tau_1) \chi_{j,x}^\beta (\tau_2) \right)^2 g_x^{\alpha\alpha'}(\tau_1)g_x^{\beta\beta'}(\tau_2)\left( \sum_j \chi_{j,x+1}^{\alpha'}(\tau_1) \chi_{j,x+1}^{\beta'}(\tau_2) \right)^2 \right) \right]
\end{align}

Next, we rewrite this fermionic partition function as a theory of bosonic bilocal fields.   Define the Green's function $G^{\alpha\beta}$ (note the replica indices):
\begin{equation}
G^{\alpha\beta}_x(\tau_1,\tau_2):= \frac{1}{N} \sum_j \chi_{j,x}^\alpha (\tau_1) \chi_{j,x}^\beta (\tau_2).
\end{equation}
We impose this definition of  $G^{\alpha\beta}_x(\tau_1,\tau_2)$  by a Lagrangian multiplier $\Sigma^{\alpha\beta}_x(\tau_1,\tau_2)$ in the path integral. Integrating out fermions after inserting $G$ and $\Sigma$ leads to the following effective theory:
\begin{align}
Z_{A,n} &= \int \calD G \calD \Sigma \exp \left( - N S[G,\Sigma] \right) \nn \\
S[G,\Sigma]& =  -\sum_x\log \Pf  \left(\partial_\tau \delta^{\alpha \beta}  - \Sigma^{\alpha \beta }_x \right) + \frac{1}{2} \sum_{x,\alpha\beta\gamma\delta} \iint_C d\tau_1 d\tau_2  \bigg[ \Sigma^{\alpha \beta}_x (\tau_1,\tau_2) G^{\alpha \beta}_x (\tau_1,\tau_2) \nn\\&
-  \frac{J_0^2}{4} G^{\alpha\beta}_x(\tau_1,\tau_2)^4
-\sum_{\delta \gamma} \frac{J_{1} ^2}{4} G^{\alpha\beta}_x(\tau_1,\tau_2)^2  G^{\gamma\delta}_{x+1}(\tau_1,\tau_2)^2 g_x^{\alpha\gamma}(\tau_1)g_{x}^{\beta\delta}(\tau_2) 
\bigg]  \label{eq:formalaction}
\end{align} 

Up to this point, all the manipulations are exact in the large $N$ limit.
In what follows, we will treat the effect of the twist operators by making certain approximations in the low temperature limit as well.

\subsection{Large $N$ and replica diagonal partition function} 

Our goal is to compute the R\'enyi entropy:
\begin{equation}
S_{A,n}= \frac{\log Z_{A,n}- n \log Z}{1-n}
\end{equation}
We have seen that $Z_{A,n}$ may be evaluated using a path integral for a generalized SYK model with $n$ flavors of replicas, with a twisted interaction on a special time contour. 
In this section, we aim to evaluate $Z_{A,n}$ with some further assumptions.

In the large $N$ limit, the partition function $Z_{A,n}$ can be computed using a saddle point approximation. We find a saddle point equation for $\Sigma$ and for $G$.   The first equation is standard:
\begin{align}
G^{\alpha\beta}_x(\tau_1,\tau_2) = ({\delta'(\tau_1,\tau_2) \delta^{\alpha\beta} - \Sigma^{\alpha\beta}_x(\tau_1,\tau_2)})^{-1}.
\end{align}
We must explicitly consider functions of  two time variables because time translation symmetry is broken due to the special time contour $C$, and the twisted interaction; the $^{-1}$ should be read as matrix inverse in both time domain $(\tau_1,\tau_2)$ and replica indices $(\alpha,\beta)$. 
The second equation depends on the location $x$.   When $x\neq 0,1$, the self energy is the same as the normal generalized SYK model (with added replica indices): 
\begin{align}
\Sigma^{\alpha\beta}_x (\tau_1,\tau_2) = J_0^2 G_{x}^{\alpha \beta}(\tau_1,\tau_2)^3+ \frac{1}{2}  J_1^2 G^{\alpha\beta}_x(\tau_1,\tau_2) \left( G^{\alpha\beta}_{x-1}(\tau_1,\tau_2)^2 + G^{\alpha\beta}_{x+1}(\tau_1,\tau_2)^2 \right)\label{SDE1}
\end{align}
Note that $(G^{\alpha\beta})^3$ means an entry-wise cube of the matrix $G^{\alpha\beta}$, and not the $\alpha\beta$ element of $G^3$. For $x=x_*$ or $x_*+1$, the self energy term experiences the twisted interaction:
\begin{align}
\Sigma^{\alpha\beta}_{x_*} &=J_0^2 (G_{x_*}^{\alpha \beta})^3+\frac{1}{2} G^{\alpha\beta}_{x_*}\left(J_1^2 (G_{x_*-1}^{\alpha\beta})^2+ \sum_{\delta \gamma} J^2_{1}g_{x_*}^{\alpha\gamma}(\tau_1)g_{x_*}^{\beta\delta}(\tau_2) ( G^{\gamma\delta}_{x_*+1} )^2 \right) \nonumber\\
\Sigma^{\alpha\beta}_{x_*+1} &=J_0^2 (G_{x_*+1}^{\alpha \beta})^3+\frac{1}{2} G^{\alpha\beta}_{x_*+1}\left(J_1^2 (G_{x_*+2}^{\alpha\beta})^2+ \sum_{\delta \gamma} J^2_{1}g_{x_*}^{\gamma\alpha}(\tau_1)g_{x_*}^{\delta\beta}(\tau_2) ( G^{\gamma\delta}_{x_*} )^2 \right) \label{SDE2}
\end{align}
where we have omitted the time variables $(\tau_1,\tau_2)$ in $G$ and $\Sigma$ for simplicity. 

Without the twist, the saddle point solution to the Schwinger-Dyson equation is diagonal in replica indices, with the form $G_x^{\alpha\beta}(\tau_1,\tau_2)=G_s(\tau_1-\tau_2)\delta^{\alpha\beta}$ \cite{stanford1604}. Since the twist couples different replicas, it is possible that the saddle point solution becomes off-diagonal. 
To see whether this possibility is realized, let us start with a theory with $J_1=0$. This reduces to a theory of decoupled SYK sites and the saddle point solution is diagonal. When a small $J_1$ is gradually turned on, it is natural to consider a perturbative solution to the Schwinger-Dyson equations (\ref{SDE1})-(\ref{SDE2}). However, according to Eq. (\ref{SDE2}) the self-energy $\Sigma^{\alpha\beta}_x$ is always proportional to $G^{\alpha\beta}_x$. Consequently, if we start from the $J_1=0$ diagonal solution and solve the equations iteratively, we find that the solution stays diagonal to {\it all orders} of the coupling $J_1^2$. Therefore we conclude that the solution either stays diagonal or that the solution is non-diagonal, but the off-diagonal contributions are non-perturbatively small as $J_1\rightarrow 0$.  In the following we will assume that $G$ and $\Sigma$ remain diagonal at finite $J_1$. 


Recall that the Renyi entropy in large $N$ limit will be determined by the saddle point with maximal contribution to the partition function. Hence, if the true minimum of the action occurs for an off-diagonal solution, then the value of $S_{A,n}$ that arises from a diagonal ansatz serves as an upper bound on the true value of $S_{A,n}$. \footnote{It should be noted that subtlety may arise from the choice of integration contour in the path integral, since only saddle points on the integration contour contributes to the partition function\cite{stanford1604}. Our argument relies on the assumption that the integration contour in $G_{x}^{\alpha\beta}$,$\Sigma_x^{\alpha\beta}$ can be chosen to include the diagonal ansatz. We thank Douglas Stanford for helpful discussion on this issue.}

With the diagonal assumption $G^{\alpha\beta}_x(\tau_1,\tau_2)=\delta^{\alpha\beta}G_x(\tau_1,\tau_2),~\Sigma^{\alpha\beta}_x(\tau_1,\tau_2)=\delta^{\alpha\beta}\Sigma(\tau_1,\tau_2)$, the effective action is simplified to
\begin{align}
\frac{1}{n} S[G,\Sigma]& = \sum_x\left\lbrace -\log \Pf  \left(\partial_\tau - \Sigma_x\right) + \frac{1}{2}  \iint_C d\tau_1 d\tau_2 \bigg[  \Sigma_x (\tau_1,\tau_2) G_x (\tau_1,\tau_2) \right. \nn\\&
\left.\left. -  \frac{J_0^2}{4} G_x(\tau_1,\tau_2)^4
- \frac{J_{1} ^2}{4} G_x(\tau_1,\tau_2)^2  G_{x+1}(\tau_1,\tau_2)^2\right]
\right\rbrace \nn \\
&+ \frac{J_1^2}{8} \left(\int_{C_1} d\tau_1 \int_{C_2}d\tau_2 + 
\int_{C_2} d\tau_1 \int_{C_1}d\tau_2
 \right) G_{x_*}(\tau_1,\tau_2)^2 G_{x_*+1}(\tau_1,\tau_2)^2
\label{eqn: effective action}\\
&\equiv \frac1nS_0[G,\Sigma]+\frac1n\Delta S[G,\Sigma]\nonumber
 \end{align} 
With the replica diagonal ansatz, the effective action is proportional to replica number $n$, so that we divide the overall $n$ to the left side. $\frac1n S_0$ denotes the first two lines, which is the original action of SYK chain, and $\frac 1n\Delta S$ denotes the third line which is the extra action cost caused by the twisted coupling. When only one of $\tau_1$ and $\tau_2$ is on the twisted contour $C_1$, the $J_1^2$ term in the action vanishes since it couples a replica diagonal term $G_{x_*}^{\alpha\alpha}$ to an off-diagonal term $G_{x_*+1}^{\alpha,\alpha+1}$ or $G_{x_*+1}^{\alpha+1,\alpha}$. This is the origin of the extra term.

\section{Weak link limit}
\label{sec: weak link limit}
Even with the replica diagonal ansatz, the Schwinger-Dyson equation is still hard to solve, especially because of the lack of time translation invariance.  However, we can start with a simple limit where
\begin{eqnarray}
N \gg \beta J \gg 1 ,~\text{and} \quad \frac{1}{\beta J} \gg \frac{J_1^2}{J^2}.
\end{eqnarray}
In this limit,  the twisted coupling term $\propto J_1^2$ can be treated perturbatively. 

To do a perturbative calculation of $S_{A,n}$, we begin by reviewing the untwisted SYK model at large $N$ and strong coupling $N\gg \beta J \gg 1$.   The saddle point solution approximately follows a conformal form:
\begin{align}
G_c(\tau_1,\tau_2)=\frac{1}{(4\pi J^2)^{1/4}} {\left( \frac{\beta}{\pi} \sin \frac{\pi (\tau_1-\tau_2)}{\beta} \right)^{-1/2}}
\label{eqn: Gc}
\end{align} 
General fluctuations around this saddle costs order $N$ action.  However, there is a special class of the fluctuations that cost action $\frac{N}{\beta J}$ in the long wave-length limit\cite{gu2016local}.  These fluctuations correspond to a time reparametrization $f_x \in \Diff(\mathrm{S}^1)$ of the conformal solution $G_c(\tau_1,\tau_2)$, which has the form \begin{equation}
G^f_x(\tau_1,\tau_2):= \left( f_x'(\tau_1) f_x'(\tau_2) \right)^\Delta G_c(f_x(\tau_1),f_x(\tau_2)), \quad \Delta=\frac{1}{4}. \label{eq:Gcreparam}
\end{equation}
In the limit $J_1^2/J^2 \ll 1/\beta J$, the twisted interaction is so weak that even these reparameterization modes will not be sourced. The first order shift to the effective action is thus obtained by evaluating the twisted term at the conformal saddle $G_c$. The effect of the reparameterization modes will be considered later in Sec.~\ref{sec: beyond 1} and \ref{sec:geo}.

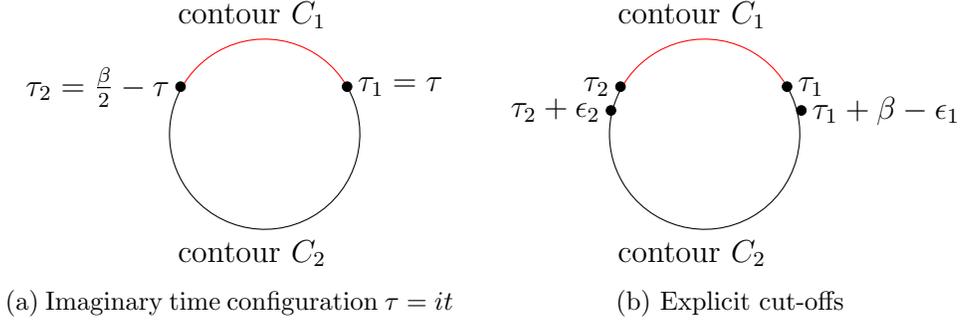
\begin{figure}[t]
\center
\subfloat[Imaginary time configuration $\tau = it$]{
\begin{tikzpicture}[scale=1.8, baseline={(current bounding box.center)}]
\draw[red] (20pt,0pt) arc (30:150:20pt);
\draw (20pt,0pt) arc (30:-210:20pt);
\filldraw (20pt,0pt) circle (1pt) node[right]{$\tau_1=\tau$};
\filldraw (-15pt,0pt) circle (1pt) node[left]{$\tau_2=\frac{\beta}{2}-\tau$};
\node at (0pt, 15pt) {contour $C_1$};
\node at (0pt, -35pt) {contour $C_2$};
\end{tikzpicture}}
\hspace{10pt}
\subfloat[Explicit cut-offs]{
\begin{tikzpicture}[scale=1.8, baseline={(current bounding box.center)}]
\draw[red] (20pt,0pt) arc (30:150:20pt);
\draw (20pt,0pt) arc (30:-210:20pt);
\filldraw (23pt,-5pt) circle (1pt) node[right]{$\tau_1+\beta-\epsilon_1$};
\filldraw (20pt,0pt) circle (1pt) node[right]{$\tau_1$};
\filldraw (-15pt,0pt) circle (1pt) node[left]{$\tau_2$};
\filldraw (-17pt,-5pt) circle (1pt) node[left]{$\tau_2+\epsilon_2$};
\node at (0pt, 15pt) {contour $C_1$};
\node at (0pt, -35pt) {contour $C_2$};
\end{tikzpicture}}
\caption{(a) The imaginary time contour involved in the entanglement entropy calculation. (b) The calculation needs to be regularized by introducing a small separation between $C_1$ and $C_2$ by $\epsilon_{1,2}$, both of which are of order $J^{-1}$.}
\label{fig: imaginary time contour}
\end{figure}

It is convenient to first work in imaginary (Euclidean) time $\tau$ and then analytically continue to real time by taking $\tau \rightarrow it$. The imaginary time problem involves a path integral with twist operators inserted at time $\tau$ and $\frac\beta 2-\tau$, as is shown in Fig.~\ref{fig: imaginary time contour}. 

At first order in $J_1^2$, the R\'enyi entropy is simply obtained by evaluating the effective action on the conformal solution $G_c $.   We find
\begin{align}
\frac{\log Z_{n,A}- n \log Z}{N} &= -\Delta S[G_c,\Sigma_c] \notag \\
&= -n \frac{J_1^2}{4} \int_{C_1} d\tau_1 \int_{C_2}d\tau_2 G_c(\tau_1,\tau_2)^4 = - n \frac{J_1^2}{8\pi J^2} \log \frac{\sin \frac{\pi}{\beta} \tau_{21}}{\sin \frac{\pi}{\beta} \epsilon}
\end{align}
where $\epsilon$ is a small UV regulator.\footnote{ Physically, this cut-off $\epsilon$ is of order $J^{-1}$: the artificial divergence arises from approximating the actual saddle point two-point function by the conformal solution $G_c(\tau)$ in Eq. (\ref{eqn: Gc}). The conformal approximation applies to the IR region $\tau \gg J^{-1}$\cite{stanford1604}, but in the UV limit $\tau\lesssim J^{-1}$ the conformal saddle diverges at $\tau\rightarrow 0$ while the true saddle $G_{\text{true}}(\tau \rightarrow 0) \rightarrow \frac{1}{2}$. The effect of this deviation can be described by a cutoff term $\log \epsilon$ with $\epsilon$ of order $J^{-1}$. }
The corresponding R\'enyi entropy is
\begin{align}
\frac{S_{A,n}}{N} = \frac{n}{n-1} \frac{J_1^2}{8 \pi J^2} \log \frac{\sin \frac{\pi}{\beta} \tau_{21}}{\sin \frac{\pi}{\beta} \epsilon}\label{Sfirstorder}
\end{align}
For convenience, we will define 
\begin{equation}
\gamma:=\frac{J_1^2}{8\pi J^2}.
\end{equation}
Analytically continuing to real time by taking $\tau_{21}= \frac{\beta}{2} - 2 it$, we obtain:
\begin{equation}
\frac{S_{A,n}}{N} = \frac{n}{n-1} \gamma \left( \log \cosh \frac{2\pi}{\beta} t - \log \sin \frac{\pi}{\beta} \epsilon  \right) 
\end{equation}
which includes a time-dependent piece and a constant piece coming from the cut-off $\epsilon$. At large real time $t\gg \beta$, the entropy grows linearly:
\begin{equation}
\frac{S_{A,n}(t)}{N} \simeq {\rm const.}+\frac{n}{n-1} \frac{2\pi \gamma}{\beta} t =\const.+ \frac{n}{n-1} \frac{J_1^2}{4 \beta J^2} t\label{RenyiPerturbative}
\end{equation}
Denote the R\'enyi entropy of each site in thermal equilibrium as $s^{\rm th}_{n}$, we can define a R\'enyi entanglement velocity as in (\ref{eq:dSEdt}): \begin{equation}
\frac{dS_{A,n}}{dt} = 2s_{A,n}^{\mathrm{th}}v_{\mathrm{E},n}
\end{equation}
The factor of $2$ comes from the fact that the TFD state is defined on two chains.
At low temperature, $s^{\rm th}_n$ approaches the finite zero temperature entropy, so that we conclude $v_{\mathrm{E},n}\propto T$ at low temperature. However, as we have discussed in the introduction (and in more detail in Sec. \ref{sec:geo}), the entanglement entropy in this system actually does not saturate to the thermal value at long time.  Thus, as we have noted previously, a conventional definition of $v_{\mathrm{E},n}$ may not apply.

Usually, taking $n\rightarrow 1$ in $S_{A,n}$ leads to the von Neumann entropy $S_A$. However, this limit is singular in Eq.~(\ref{RenyiPerturbative}).   This is a consequence of the replica diagonal ansatz, because the resulting effective action $S\propto n$.  Physically, the divergence suggests that the $n-1\ll 1$ region is described by a replica non-diagonal saddle point. But we emphasize that $S_{A,n}(t)$, as given in (\ref{Sfirstorder}), is an upper bound for the R\'enyi entropy of the optimal non-diagonal solution. We will return to this point in Sec.~\ref{sec:results}.

\section{Deviation from linear growth: Gaussian correction}
\label{sec: beyond 1}

For any chain with a finite length, entropy is upper bounded.   The linear growth of entropy cannot last forever.  To see the saturation of $S_{A,n}(t)$ as $t$ becomes large, we need to go beyond the first order approximation of the previous section.
In this section we analyze the second order correction to the first order linear growth.  More explicitly, we will consider the change of the saddle point solution due to the additional twisted interaction term.
Recall that this amounts to finding the minimum of the effective action expressed in Eq.~(\ref{eqn: effective action}):
\begin{equation}
\log Z_{A,n} = - \min_{\{ G,\Sigma \}} S[G,\Sigma]
\end{equation}
As we noted before in Eq. (\ref{eq:Gcreparam}), so long as  $\frac1{\beta J}\ll 1$, the soft modes of the SYK model are reparameterizations of the untwisted saddle point solution.
When the twisted coupling is also small:  $\gamma \ll 1$, we can ignore the induced change of $G$ outside the manifold of reparameterization. In this approximation, the saddle point solution is determined by minimizing the effective action $-\log Z_{A,n}[G^f_x(\tau_1,\tau_2)]$ over reparameterization $f(\tau)$. Since the minimal action in the restricted space of $G_x^f(\tau_1,\tau_2)$ is always larger or equal to the actual minimal action in the unrestricted space of two-point functions, the entropy we obtain in this approximation always bounds the actual entropy from above. In other words, our results are still meaningful as an upper bound, even when the effect of non-reparameterization modes is not negligible.

\subsection{Effective action of the reparameterization field}

The form of the effective action $S[f_x(\tau)]\equiv -\log Z_{A,n}\left[G^f_x\right]$ can be explicitly written down. For simplicity, we will consider a chain with an even number $M\in 2\mathbb{Z}$ of sites $x=-\frac{M}{2}+1,-\frac{M}{2}+2,\ldots, \frac{M}{2}-1,\frac{M}{2}$.We choose open boundary conditions, and an entanglement cut in the middle, between sites $x_*=0$ and $x_*+1=1$. In this case, the system has reflection symmetry after the random average, so that the saddle point solution shall satisfy $f_{x}(\tau)=f_{-x+1}(\tau)$. In particular, $f_0(\tau)=f_{1}(\tau)$. 
With this simplification one can write
\begin{align}
S&=S_0+\Delta S\nonumber\\
\frac1 nS_0&=
\sum_x \left[ - \alpha_{\mathrm{S}} \int d \tau \Sch\left(e^{i \frac{2\pi}{\beta} f_x}, \tau \right) + \frac{J_1^2}{16} 
\int d\tau^2 \left( (G^f_{x})^2-(G^f_{{x+1}})^2\right)^2
 \right]
\\
\frac1 n\Delta S&=\frac{J_1^2}{4} \int_{C_1} d\tau	_1 \int_{C_2} d\tau_2 G^f_{{x_*}}(\tau_1,\tau_2)^4
\end{align}
The $S_0$ term controls the dynamics of the reparametrization field in SYK chain model\cite{gu2016local} without the twisted interaction;.
$\alpha_{\mathrm{S}}\approx 0.01$ is the numerical coefficient of the Schwarzian term\cite{stanford1604}, and this coefficient also determines the specific heat: $c_v= N \frac{8\pi^2 \alpha_{\mathrm{S}}}{J}$.\footnote{This specific heat is for the TFD state, which is twice of the specific heat for a single chain. \cite{gu2016local}} 
Without the twist term, the SYK chain model has a saddle point solution $f(\tau)=\tau$, which corresponds to the conformal solution $G_c$.
Using the explicit form of $G^f_{0}$ in terms of the conformal saddle $G_c$, we can further write
\begin{eqnarray}
\frac1 n\Delta S=\frac{J_1^2}{4}\int_{f(C_1)} df(\tau_1) \int_{f(C_2)} df(\tau_2) G_c(f(\tau_1),f(\tau_2))^4
\end{eqnarray}
This integral diverges due to the UV divergence of $G_c$ at $\tau_1-\tau_2\rightarrow 0$. This divergence is artificial, since the actual saddle point two-point function should saturate to the UV value $\frac12$ for $\tau_1 - \tau_2 \lesssim J^{-1}$. To take into account of this UV regularization, we introduce UV cut-off for $C_1$ and $C_2$.  On the imaginary time circle, this corresponds to defining $C_1$ and $C_2$ as interval $[\tau_1,\tau_2]$ and $[\tau_2+\epsilon_2,\beta+\tau_1-\epsilon_1]$, respectively, 
as shown in Fig.~\ref{fig: imaginary time contour}(b). The resulting integral can be evaluated explicitly, using the functional form of $G_c$ given in Eq.~(\ref{eqn: Gc})
\footnote{The evaluation essentially corresponds to the explicit integral
\begin{align*}
I&= \int_{\tau_1}^{\tau_2} d\tau  \int_{\tau_2+\epsilon_2}^{\beta+\tau_1-\epsilon_1} d\tau' \frac{1}{\left( \sin \frac{\pi (\tau-\tau')}{\beta} \right)^2} 
= \left( \frac{\beta}{\pi} \right)^2\log  \frac{\sin \frac{\pi}{\beta} (\tau_2-\tau_1+\epsilon_1)}{\sin \frac{\pi}{\beta} \epsilon_1} \cdot
\frac{\sin \frac{\pi}{\beta} (\tau_2-\tau_1+\epsilon_2)}{\sin \frac{\pi}{\beta} \epsilon_2} .
\end{align*}
This is simply the cross ratio (\ref{eq:crossratio}) for four points on the imaginary time circle, for $f_0(\tau)=\tau$.
} and the result is
\begin{equation}
\frac{1}{n} \Delta S= \frac{\gamma}{2} \cdot  \log \eta[f_{x_*}],
\end{equation}
where $\eta[f_{x_*}]$ is the reparametrized cross ratio:
\begin{align}
\eta[f_{x_*}]&= \frac{\sin \frac{\pi}{\beta} (f_{x_*}(\tau_2)-f_{x_*}(\tau_1-\epsilon_1))}{\sin \frac{\pi}{\beta} (f_{x_*}(\tau_1+\epsilon_1)-f_{x_*}(\tau_1))} \cdot
\frac{\sin \frac{\pi}{\beta} (f_{x_*}(\tau_2+\epsilon_2)-f_{x_*}(\tau_1))}{\sin \frac{\pi}{\beta} (f_{x_*}(\tau_2+\epsilon_2)-f_{x_*}(\tau_2))}. \label{eq:crossratio}
\end{align}
As expected, the result is explicitly invariant under the global conformal group $\SL(2,\RR)$.
Therefore we have computed the twisted interactions, restricted to the reparameterization modes $f_x(\tau)$.

In general, it is still difficult to rigorously study the effective action of reparameterization fields. In the remainder of this section we will study the effective action and R\'enyi entropy using a Gaussian approximation.    In the next section we will study the full non-linear effective action, but mostly for the simplified case of two-site problem.

\subsection{Gaussian correction}\label{sec:gaussian}
In the spirit of our previous perturbative calculation as $\gamma \rightarrow 0$, we now compute the second order correction to $S_{A,n}$ in $\gamma$. 
Defining $f_x(\tau)=\tau+\epsilon_x(\tau)$, we must expand the action up to the quadratic order in $\epsilon(\tau)$, compute the first order correction $\epsilon(\tau) \sim \gamma$ caused by the twisted interaction term in the action, and then evaluate the action to order $\gamma^2$, accounting for the non-zero $\epsilon(\tau) $.   The effective action for small $\epsilon(\tau) $ is given by
\begin{eqnarray}
\frac1n S[\epsilon]&\simeq &\frac{\alpha_{\mathrm{S}}}{\beta J} \sum_{n,p} \frac{1}{2} |n|\left(|n| + \frac{\beta}{2\pi} D p^2\right)(n^2-1) \epsilon_{n,p} \epsilon_{-n,-p} \nonumber\\
& &+\frac\gamma 2\left[\log \eta_0+\frac{2\pi}{\beta} \left( 
\frac{\epsilon_{x_*}(\tau_1)-\epsilon_{x_*}(\tau_2)}{\tan \frac{\pi}{\beta} \tau_{12}}  - \frac{\beta}{2\pi} ( \epsilon'_{x_*}(\tau_1) + \epsilon'_{x_*}(\tau_2) )
\right)
\right]
\end{eqnarray}
The first line is the quadratic expansion of $S_0$, which has been obtained in \cite{gu2016local}. The second line corresponds to $\frac1n\Delta S$. $\eta_0=\eta[f(\tau)=\tau]$ is the cross ratio for the trivial reparameterization, which corresponds to the first order contribution to the entropy we obtained earlier in Eq. (\ref{Sfirstorder}).  The twist term contributes a linear in $\epsilon$ term, which directly sources  the first order correction $\epsilon(\tau) \sim \gamma $.   We have ignored the quadratic term in $\Delta S$;  it will modify $S_{A,n}$ at higher order in $\gamma$.  Minimizing $S[\epsilon]$ is straightforward:
\begin{align}
\frac{1}{n} \min_{\epsilon}  S[\epsilon] =  \frac{\gamma}{2} \log \eta_0  -  \frac{1}{2} \left\langle  \left( \frac{\Delta S}{n} \right)^2 \right\rangle,
\end{align}
where $\langle \cdots \rangle$ denotes expectation values with respect to the quadratic action for $\epsilon(\tau)$ given above.
In Appendix~\ref{appendix: infinite chain}, we explicitly perform this expectation value, and we find that
\begin{align}
 \frac{1}{n} \min_{\epsilon}  S[\epsilon] \simeq \frac{2\pi \gamma t}{\beta}  \cdot  \left( 1- \frac{2}{\sqrt{3} \pi^2 } 
\sqrt{\frac{ J \gamma t}{\alpha_{\mathrm{S}}} }
   \right)
\end{align}

The negative correction $\propto t^{3/2}$ indicates that the entropy growth starts to become slower than linear 
around the characteristic time $t_* \sim \frac{\alpha_{\mathrm{S}}}{\gamma J}$. 
In the weak link limit $\gamma \ll \frac{1}{\beta J}$  discussed here, this time scale $t_*\sim \beta \frac{1}{\gamma \beta J} \gg \beta$ is much longer than the thermal time $\beta$. We can further estimate the amount of entropy growth by the time $t_*$:
\begin{align}
\Delta S_{A,n}=S_{A,n}(t_*)-S_{A,n}(0) \sim N \frac{n}{n-1} \cdot \frac{2\pi \gamma}{\beta} t_* \sim \frac{n}{n-1} \frac{N\alpha_{\mathrm{S}}}{\beta J}
\end{align}
which is of order $c_v T$, the specific heat's contribution to the thermal entropy. As we will see later in Sec.~\ref{sec:geo}, the entropy actually saturates to a final value that is comparable with our estimation here.

\section{Long time saturation: Geometric interpretation}\label{sec:geo}

In this section, we will evaluate the partition function $Z_{A,n}$ for a simple case, when the chain has only two coupled SYK sites ($M=2$).  The entanglement cut is between the two sites.   For the two-site problem, the full non-linear saddle point problem can be solved using a geometric interpretation of the action. Although this special case does not directly determine the entropy growth in a longer chain, the results will help us to understand qualitative features of this system, especially the long time saturation of the (R\'enyi) entanglement entropy.\footnote{Since we have taken the large $N$ limit first, even for two sites it is not immediately clear that entropy growth will stop at late time.} As we will discuss in Sec.~\ref{sec:results}, the two-site calculation can be generalized to longer chains, and by doing so, we provide an upper bound of the entropy growth and saturation in that case. 

\subsection{Two site effective action and the mapping to a geometric problem}

As we discussed in the previous section, the two-site problem with an entanglement cut between the two sites has a reflection symmetry, so that the saddle point solution should be given by identical reparameterization fields on the two sites: $f_1(\tau)=f_2(\tau)=f(\tau)$. The effective action is thus a functional of a single $f(\tau)$, 
and the R\'enyi entropy has the following form:
\begin{align}
S_{A,n}
= \frac{n}{n-1} N \min_{f} \left(  - \frac{2\alpha_{\mathrm{S}}}{J} \int d\tau \Sch\left(\tan \frac{\pi}{\beta} f(\tau) , \tau \right) + \frac{\gamma}{2} \log \eta_f + 2\alpha_{\mathrm{S}} \frac{2\pi^2}{\beta J}   \right)\label{eq: entropy for repara}
\end{align}
The first term is the Schwarzian action for the reparametrizations $f$ on two sites (hence the factor of 2), and 
the second term arises from the twisted interaction between the two sites. The last term, coming from the $\log Z$ term, is the constant piece of the Schwarzian, which cancels the  value of the first term when $f(\tau)=\tau$. Therefore, our goal here is to find the minimal value of $S_{A,n}$ by varying all reparametrizations $f \in \Diff(S^1)$:
\begin{align}
I(t) = \min_f  \left(  - \frac{2\alpha_{\mathrm{S}}}{J} \int d\tau \Sch\left(\tan \frac{\pi}{\beta} f(\tau) , \tau \right) + \frac{\gamma}{2} \log \eta_f  \right)\label{two site action}
\end{align}
The time dependence comes from the second term $\log \eta_f$
, where $\eta_f$ is the cross ratio of the reparametrization of four time coordinates: $(\tau_1,\tau_2,\tau_2+\epsilon_2,\beta+\tau_1-\epsilon_1)
$ (cf. (\ref{eq:crossratio})) and $\epsilon_{1,2}$ are cut-offs of order $J^{-1}$. $\tau_1$ and $\tau_2$ will be analytic continued to $it$ and $\frac{\beta}{2}-it$ towards the end of the calculation. In the limit $\epsilon_{1,2}/\beta \sim (\beta J)^{-1}\ll 1$, $\eta_f$ is simplified to 
\begin{align}
\eta_f & \simeq \frac{\sin \frac{\pi}{\beta} (f(\tau_2)-f(\tau_1))^2}{(\frac{\pi}{\beta})^2 \epsilon_1 \epsilon_2 f'(\tau_1) f'(\tau_2)} 
\end{align}
It is manifest that both terms in the two-site action (\ref{two site action}) are $\SL(2,\RR)$ invariant.

The Schwarzian action term has a geometric interpretation \cite{maldacena2016conformal,kitaevIAS}, which corresponds to the area enclosed by a curve in hyperbolic space with fixed length
 \begin{equation}
L=\beta J \gg 1
\end{equation}
More explicitly, one can consider a hyperbolic disk with coordinates $(\rho,\theta)$ and metric:
\begin{align}
ds^2 = d\rho^2 +\sinh^2 \rho d\theta^2
\label{eq:globalads2metric}
\end{align}
We specify a curve on the disk by $(\rho(\tau),\theta(\tau))$. For each reparametrization $f(\tau)$, we let $\theta(\tau) = \frac{2\pi}{\beta} f(\tau)$ and $\rho(\tau)$ determined by the constrain that induced metric $g_{\tau \tau}(\tau)= J^2$ along the curve is fixed. Then one can show (see Appendix~\ref{appendix: Geometric interpretation} for details) in the large $L=\beta J$ limit, the Schwarzian action:
\begin{equation}
 - \frac{\alpha_{\mathrm{S}}}{J} \int d\tau \Sch\left(\tan \frac{\pi}{\beta} f(\tau) , \tau \right)  \simeq  \alpha_{\mathrm{S}} (L-A-2\pi).
\end{equation}
Here $A$ is the area enclosed by the curve.

\begin{figure}[t]
\center
\subfloat[Generic curve with fixed length]{
\begin{tikzpicture}[scale=0.8,baseline={(current bounding box.center)}]
\draw[thick] (0pt, 0pt) circle (100pt);
\filldraw (40pt,40pt) circle (1pt) node[above right]{$X_1$};
\filldraw (-40pt,-60pt) circle (1pt)  node[left]{$X_2$};
\draw[fill=yellow, opacity=0.40] (40pt,40pt)  .. controls (120pt, 30pt) and (60pt,-100pt) .. (-10pt,-85pt) .. controls (-40pt,-80pt) and (-40pt, -70pt)..
(-40pt,-60pt) .. controls  (-70pt,-40pt) and (-90pt, 40pt)..(-40pt, 60pt) ..controls  (-10pt, 75pt) and (10pt, 70pt)..(40pt,40pt) ;
\draw[dashed, thick] (40pt,40pt) -- (-40pt,-60pt);
\end{tikzpicture}
}
\hspace{30pt}
\subfloat[Piece-wise arcs with fixed total length]{
\begin{tikzpicture}[scale=0.8,baseline={(current bounding box.center)}]
\draw[thick] (0pt, 0pt) circle (100pt);
\draw[dashed] (0pt, 0pt) -- (100pt,0pt);
\filldraw (60pt,40pt) circle (1pt) node[above right]{$X_2$} ;
\filldraw (60pt,-40pt) circle (1pt) node [below right]{$X_1$};
\node at (-60pt,60pt) {$C_2$};
\node at (90pt,20pt) {$C_1$};
\filldraw (0pt,0pt) circle (1pt) node [below] {$O$};
\draw [black,dashed] (0pt,0pt) -- (60pt, 40pt);
\draw [black,dashed] (0pt,0pt) -- (60pt, -40pt);
\draw[red, dotted] (60pt,40pt) arc (63.43: 296.57 : 44.7pt);
\draw[thick, red] (60pt,40pt) arc (63.43: -63.43 : 44.7pt);

\draw[thick, blue] (60pt,40pt) arc (33.69: 326.31 : 72pt);
\draw[dotted, blue] (60pt,40pt) arc (33.69: -33.69 : 72pt);
\filldraw (40pt,0pt) circle (1pt) node [below] {$C$};
\draw[dashed] (40pt,0pt) -- (60pt,40pt);
\draw[black, dashed](60pt,40pt) -- (60pt, -40pt);
\end{tikzpicture}}
\caption{
(a) Geometric illustration of the minimization problem we need to solve. The total length of the curve is fixed to be $L$. The ``string'' connects $X_1$ and $X_2$ tends to pull the two points closer, while the other term of the action prefers the curve to enclose a larger area.
(b) For a fixed distance $D(X_1,X_2)$ between $X_1$ and $X_2$, the joint of two arcs maximizes the area enclosed by the curve. Then the entropy is determined by the minimum of action as a function of $D(X_1,X_2)$. 
}
\label{fig: A of D}
\end{figure}
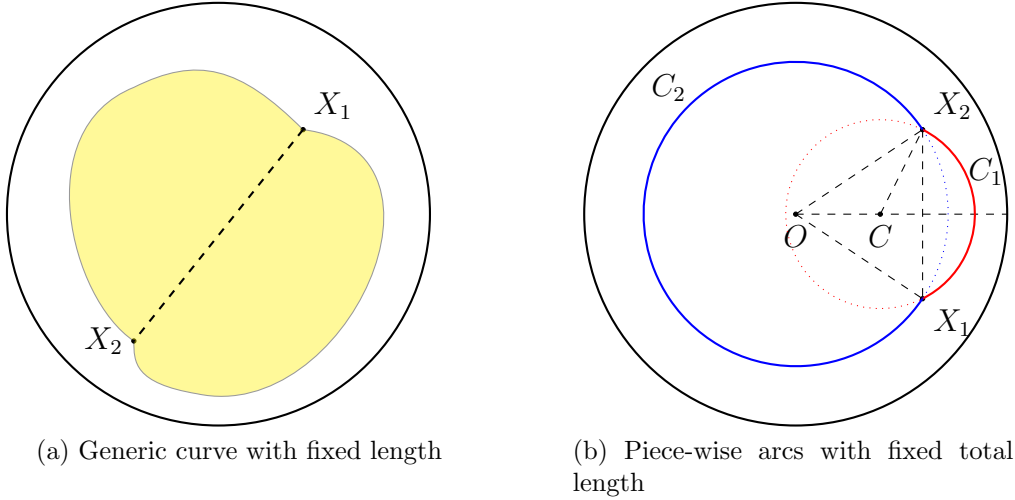

The second term of the two site action (\ref{two site action}) also has a simple geometric interpretation in the embedded picture. 
In the strong coupling limit $L=\beta J \gg 1$ we are interested, the curve is very close to the boundary and the cross ratio term can be written as:
\begin{equation}
\log \eta_f =  \log \cosh D(X_1,X_2) 
\end{equation}
where $X_{1,2}$ is the point on the curve that corresponds to the twist operator time $\tau_{1,2}$, respectively. $D$ is the distance function with respect to the metric (\ref{eq:globalads2metric}). For more details see Appendix~\ref{appendix: Geometric interpretation}.
Therefore, the minimization problem for $I(t)$ is, in fact, a simple geometric problem:
\begin{align}
I(t) = \min \left( 2 \alpha_{\mathrm{S}} (L-A-2\pi)+ 
 \frac{\gamma}{2} \log \cosh D(X_1,X_2)
\right) \label{geo action}
\end{align}
See Fig.~\ref{fig: A of D}(a) for an illustration. The first term in $I(t)$ prefers to maximize the area $A$ enclosed by the curve, for fixed length $L$. In the untwisted problem (without the second term), the maximal area is obtained by a circle, corresponding to $f(\tau)=\tau$ (up to an $\SL(2,\RR)$ transformation).  The second term adds an ``attractive force" between two particular points on the curve. 
The locations of these two points are  fixed by demanding an arc length $J\tau_1=J\tau$ and $J\tau_2=\frac{\beta J}{2}-J\tau$, starting from a reference point.\footnote{The location of the reference point is not important since it can be moved around by an $\SL(2,\RR)$ transformation.}  In the $L\rightarrow \infty$ limit, the distance between the two points is large.  The attractive force $\propto \log \cosh D(X_1,X_2)\simeq D(X_1,X_2)$ is a confining force that grows linearly with distance. The intersite coupling $\gamma=\frac{J_1^2}{8\pi J^2}$ plays the role of a ``string tension" connecting the two points. The competition of these two terms determine the saddle point solution $f(\tau)$.

The geometric interpretation greatly simplifies the problem. Since the second term only depends on the position of two points $X_1,X_2$ on the curve, the minimization of action $I(t)$ can be considered as a two-step process. Firstly, we maximize the area $A$ for a given position of $X_1,X_2$. Secondly, we change the distance between $X_1,X_2$ to minimize the action. In the first step, when $X_1,X_2$ are fixed, we are varying two curves, one with the fixed length of $\tau_2-\tau_1=\frac\beta 2-2\tau$ and the other with fixed length $\frac{\beta}2+2\tau$. Obviously, maximizing the area requires each of them to be an arc, as shown in Fig.~\ref{fig: A of D}(b). For a given distance $D(X_1,X_2)$ between the two points, the shape of each arc is completely determined, so that the minimized action $I(t)=\min_D I(D)$ is now determined by one parameter $D=D(X_1,X_2)$. The second step is thus a one-parameter minimization problem which can be easily studied analytically and numerically.

In principle, the geometric minimization problem can be worked out  for arbitrary $L$ and $\gamma$. However, the  geometric interpretation that relates the Eq.~(\ref{two site action}) to Eq.~(\ref{geo action}) is only valid if the curve in the hyperbolic disk is close to the conformal boundary. Beyond this limit, the Schwarzian action and the area term are not equivalent, and there are subleading 
corrections which must be accounted for. 
A sufficient condition for the curve to be close to the boundary is $\gamma \lesssim \frac1{L}\ll 1$. In this limit the ``string tension" $\gamma$ is sufficiently weak such that the curve is not deformed too much.

\subsection{Explicit form of the action}
We now follow the procedure outlined above and write down the action $I(D)$ for imaginary time explicitly as a function of $D$.
Denote the radius of the two arcs ($C_1$ and $C_2$ in Fig.~\ref{fig: A of D}(b))as $\rho_{1,2}$, and the `opening angles' as $\alpha_1$ and $\alpha_2$. These four parameters are determined by the arc length constraint and the distance $D=D(X_1,X_2)$. By construction, the arc length of $C_1$ and $C_2$ are fixed, which implies
\begin{align}
\alpha_1 \sinh \rho_1 = Lx, \quad
\alpha_2 \sinh \rho_2 = L(1-x) \label{geoEq1}
\end{align}
with $x=\frac12-\frac{2\tau}{\beta}$. In addition, the distance between the end points $X_1,X_2$ satisfies
\begin{align}
\cosh D= 1+ 2 \sinh ^2 \rho_1 \sin ^2 \frac{\alpha_1}{2}=1+ 2 \sinh ^2 \rho_2 \sin ^2 \frac{\alpha_2}{2}\label{geoEq2}
\end{align}
These four equations determine all four parameters $\rho_{1,2},\alpha_{1,2}$ as functions of $D$. 

The area enclosed by the joint of the two arcs can be divided into four parts, including two circular wedges and two triangles.
The area of the wedges are easy to compute:
\begin{align}
A_{\text{wedges}} = (\cosh \rho_1-1) \alpha_1 +(\cosh \rho_2 -1) \alpha_2
\end{align}
We can then use the Gauss-Bonnet theorem to compute the area of triangles. As all extrinsic curvature of the triangular regions is located at the corners, we simply sum the inner angles.
Two of the inner angles are known.   The nontrivial inner angle is $\angle CXO:=\phi$ in Fig.~\ref{fig: A of D}(b) which can be computed by considering the angle between radial line $OX/CX$ and geodesic $X_1 X_2$:
\begin{align}
\phi_1 = \arctan \left(  - \frac{\cot \frac{\alpha_1}{2}}{\cosh \rho_1} \right) , \quad 
\phi_2 = \arctan \left(  \frac{\cot \frac{\alpha_2}{2}}{\cosh \rho_2} \right) , \quad
\phi= \phi_1 - \phi_2
\end{align}
Therefore, the total area of the two triangles is $
A_{\vartriangle}= \alpha_1+\alpha_2 - 2\phi - 2\pi
$.
Evaluating each part with hyperbolic geometry, we obtain 
\begin{align}
A=\cosh \rho_1 \alpha_1 +\cosh \rho_2 \alpha_2 +2\arctan\left(\frac{\cot\frac{\alpha_1}2}{\cosh\rho_1}\right)+2\arctan\left(\frac{\cot\frac{\alpha_2}2}{\cosh\rho_2}\right) - 2\pi\label{geoArea}
\end{align}
Using Eqs. (\ref{geoEq1}), (\ref{geoEq2}), and (\ref{geoArea}), we can express the action $I(D)$ as a function of $D$.   $I(t)$ will be the minimal value of $I(D)$ when varying $D$. 

\subsection{Analytic continuation}
The imaginary time minimization problem can be solved numerically, which leads to $I(\tau)$ as a function of imaginary time $\tau$. However, only knowing $I(\tau)$ numerically makes the analytic continuation difficult. Instead, we address the real time problem directly, and analytically continue the equations (\ref{geoEq1}), (\ref{geoEq2}) and (\ref{geoArea}).  The analytic continuation is defined by $\frac{2\tau}{\beta}\rightarrow\frac{i2t}{\beta}$, or $x=\frac12-i\frac{2t}{\beta}$ in Eqs. (\ref{geoEq1}) and (\ref{geoEq2}). This leads to complex $\alpha_{1,2}$ and $\rho_{1,2}$.    
In general, this would lead to a complex action (and thus complex entropy), which would be unphysical.   However, we notice that 
$\alpha_1\sinh\rho_1=L(\frac12-i\frac{2t}{\beta})$ and $\alpha_2\sinh\rho_2=L(\frac12+i\frac{2t}{\beta})$ are complex conjugates. Both Eqs. (\ref{geoEq1}) and (\ref{geoEq2}) have a $\mathbb{Z}_2$ symmetry: 
\begin{align}
\alpha_1\leftrightarrow \alpha_2^*, ~\rho_1\leftrightarrow \rho_2^*,~D\leftrightarrow D^*
\end{align}
Thus we can look for saddle point solution that is invariant under this $\mathbb{Z}_2$ symmetry, which satisfies $\alpha_2=\alpha_1^*,~\rho_2=\rho_1^*,~D\in\mathbb{R}$. In this case Eqs. (\ref{geoEq1}) and (\ref{geoEq2}) require
\begin{align}
\sinh\frac{D}2=\frac{L\left(\frac12-i\frac{2t}{\beta}\right)}{\alpha_1}\sin\frac{\alpha_1}2\in\mathbb{R},\label{Drealitycondition}
\end{align}
and the area term becomes
\begin{align}
A=2{\rm Re}\left[\sqrt{L^2\left(\frac12-i\frac{2t}{\beta}\right)^2+\alpha_1^2}+2\arctan \left( \frac{\cot \frac{\alpha_1}{2}}{ \sqrt{\left( \frac{L(\frac12-i\frac{2t}{\beta})}{\alpha_1} \right)^2+1 }} \right) \right]-2\pi \label{eq:realarea}
\end{align}
The complex angle $\alpha_1$ is determined by $D$ (which remains real) in Eq.~(\ref{Drealitycondition}). The resulting action $I(D)$ is the sum of (\ref{eq:realarea}) and $\frac{\gamma}{2}\log\cosh D$, and both terms are manifestly  real. We can now minimize $I(D)$ with respect to $D$ directly for real time $t$. 

\subsection{Numerical and analytic results}\label{sec:results}

\begin{figure}[t]
\includegraphics[width=8cm]{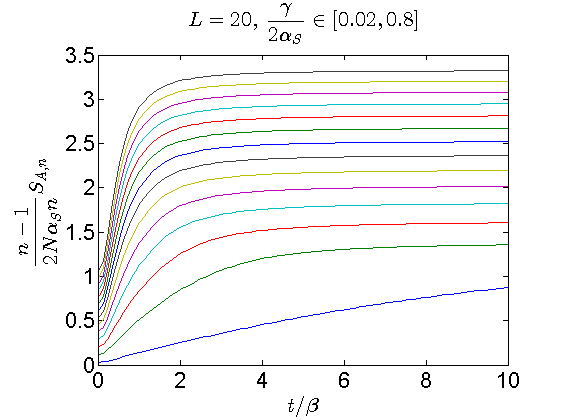}\includegraphics[width=8cm]{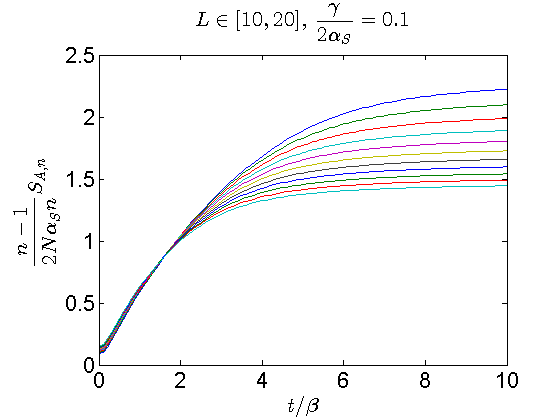}
\centering
\caption{Numerical results of the entropy as a function of time and parameters $L, \gamma$. The entropy is measured in unit $\frac{2N\alpha_Sn}{n-1}$ (the coefficient of the area term in the expression of entropy (\ref{eq: entropy for repara})). Left panel shows the entropy growth for fixed $L=20$ and different $\gamma$ (higher $\gamma$ corresponds to curves with higher entropy). Right panel shows the entropy growth for different $L\in[10,20]$ for fixed $\frac{\gamma}{2\alpha_S}=0.1$.\label{fig:numerics}}
\end{figure}

With the analytically continuated effective action defined above, we can minimize the action with respect to $D$ numerically, and obtain the R\'enyi entropy $S_{A,n}(t)$. 
The numerical result is shown in Fig.~\ref{fig:numerics}. 
We see that the entropy grows quadratically at very early time, and subsequently crosses over to a linear growth. The linear growth rate agrees with the perturbative calculation for small $\gamma$. At longer times, the growth slows down and eventually $S_{A,n}$ saturates to a finite value as $t\rightarrow \infty$. Therefore the result appears to be qualitatively similar to the expectation for a thermalizing system.
However, as we will show, this system has not actually thermalized.

To gain better understanding of the long time saturation behavior, we have observed numerically that at long time the saddle point value of $D$ corresponds to $\Re \alpha_1\rightarrow 2\pi$. Therefore we can expand around this point and study the long time behavior. In this limit, Eq. (\ref{Drealitycondition}) requires $\alpha_1$ to take the following form (for details, see Appendix~\ref{sec: constrain}):
\begin{align}
\alpha_1= 2\pi-\frac{\delta^2}{2\pi} + \frac{\delta \beta}{4t}+i\delta
\end{align}
with $\delta\in\mathbb{R}$. In the limit $\delta\ll 1,~L\gg 1,~\gamma L \ll 1$, the saddle point value is $\delta_*=-\frac{16 \alpha_{\mathrm{S}} \pi \beta}{\gamma Lt}$, which corresponds to the entropy 
\begin{align}
\frac{n-1}{n N}S_{A,n}(t\rightarrow +\infty)&\simeq \frac{n-1}{n N}   S_{A,n}(\infty)-c\frac{\beta^2}{t^2} + \ldots 
\end{align}
where the saturation value is approximately given by
\begin{align}
\frac{n-1}{n N}S_{A,n}(\infty)&\simeq  \frac{4\pi^2\alpha_{\mathrm{S}}}{\beta J}+\gamma\left(1+\log\frac{8\sqrt{2}\alpha_{\mathrm{S}}}{\gamma} \right)   \label{longtime limit}
\end{align}
and the coefficient for the saturation term is given by \begin{equation}
c\simeq 
\frac{\alpha_{\mathrm{S}} }{L} \left( 
\frac{128 \pi^2 \alpha_{\mathrm{S}}  }{3 \gamma L } + \pi^2+2
 \right). \label{eq:cinmain}
 \end{equation}
Details of this calculation are presented in  Appendix~\ref{appendix: value of c}.
It should be noticed that the first term is $\gamma$-independent, which clearly shows that the result is non-perturbative in $\gamma$ (as we have first taken $t\rightarrow \infty$), although we still consider a small $\gamma \ll L^{-1}$.

The entropy at $t=0$ can be computed perturbatively, leading to
\begin{align}
\frac{n-1}{nN} S_{A,n}(0)&\simeq \frac\gamma{2}\log\frac{L^2}{2\pi^2}
\end{align}
In the limit where $t\rightarrow \infty$, and then $\gamma\rightarrow 0$, the entropy growth is thus given by
\begin{align}
\Delta S_{A,n}\frac{n-1}n&\simeq \frac{4 N\pi^2  \alpha_{\mathrm{S}}}{\beta J}= \frac{1}{2} c_vT\label{entropy growth}
\end{align}
with $c_v$ the specific heat of the doubled SYK model we are studying. The entropy growth is independent from the UV cut-off. The relation to specific heat is interesting, since $c_vT=S_{\rm th}(T)-S_{\rm th}(0)$ is the thermal entropy minus the zero temperature entropy. Therefore, even after a long time, the TFD state has not thermalized, and the entropy is smaller than the thermal entropy by an amount that is determined by the zero temperature extremal entropy.

The result of our geometric minimization procedure can also be verified by applying the Gaussian approximation method in Sec.~\ref{sec: beyond 1} to the two-site problem. 
We find that the early time linear growth will be corrected by a quadratic term around time scale $t_*\sim \frac{\alpha_{\mathrm{S}}}{\gamma L}$ in the weak link limit $\gamma L \ll 1$:
\begin{align}
\Delta S_{A,n} \frac{n-1}{n} \simeq \frac{2\pi \gamma t}{\beta} N \left( 1-  \frac{\gamma J t}{8 \pi \alpha_{\mathrm{S}}}  \right)
\end{align}
In Appendix~\ref{appendix: two sites gaussian}, we present the detail of the Gaussian approximation calculation, and compare the result with the geometric formula in Appendix~\ref{appendix: two sites geo}. 
This early time result suggests that the $t^2$ term becomes significant when the entropy approaches $\propto \frac{N\alpha_{\mathrm{S}}}{\beta J}$, which is consistent with the late time saturation value we get above.

\begin{figure}[t]
\includegraphics[width=8cm]{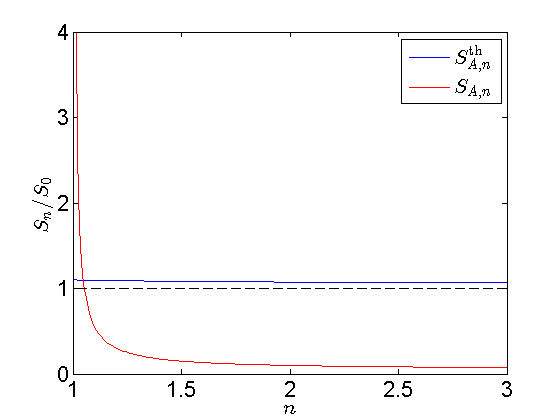}
\centering
\caption{Comparison of the R\'enyi entropy $S_{A,n}$ that we obtained in Eq. (\ref{entropy growth}) and the thermal R\'enyi entropy. The entropies are measured by the zero temperature entropy, and the curves are plotted for $\frac{c_vT}{S_0}=0.1$.  \label{fig:thermal}}
\end{figure}

Thus, we conclude that the R\'enyi entanglement entropy in the time-evolved TFD state saturates to a sub-thermal value, and in the low temperature and weak inter-site coupling limit, the entropy is proportional to the ``near-extremal entropy"
$S_{A,n}^{\rm th}(T)-S_{A,n}^{\rm th}(T=0)$. 
This result indicates that the system does not completely thermalize, but instead reaches a ``pre-thermalized" state. Roughly speaking,  pre-thermalization occurs because the degrees of freedom in this system are separated into fast modes (the reparameterization quasi-Goldstone modes) and slow modes (which are responsible for the zero temperature entropy). The latter are almost localized, and so it is natural that they require a long time to thermalize.  The lack of rapid thermalization for these slow modes seems consistent with the fact that the non-reparameterization modes in the coupled SYK chain have exponentially decaying correlation functions, with correlation length at the order of lattice constant\cite{gu2016local}. The thermalization time may grow with some function of $N$ but must diverge in the infinite $N$ limit.
We expect that at finite $N$, the thermalization time for the SYK chain is finite, as it seems unlikely that the slow modes of the SYK chain have many-body localized.

Our result is based on several assumptions. We have taken a replica diagonal ansatz of the two-point function, and then further restricted ourselves to the reparameterization of the unperturbed saddle point. The entropy we obtain is determined by minimization of the effective action in this restricted space. 
As we have previously noted, the actual entropy obtained by unrestricted minimization will be smaller or equal to what we obtained. Thus, the pre-thermalization feature we have uncovered is independent of the validity of our approximations, and remains true even for the ``actual" saddle point solution $G^{\alpha\beta}_x(\tau_1,\tau_2)$. 

We have computed the R\'enyi entanglement entropy of the SYK chain in a TFD state, whereas most previous studies of similar systems have studied the von Neumann entropy.  As such, we now discuss the analytic continuation of the R\'enyi entropy to $n\rightarrow 1$.  We can compare the long time entropy  (\ref{entropy growth}) with the R\'enyi entropy of the thermal ensemble for the same system $A$, which is
\begin{align}
S_{A,n}^{\rm th}=S_0+\left(1+\frac1n\right)\frac12c_vT
\end{align}
The comparison of $S_{A,n}^{\rm th}$ and $S_{A,n}(t\rightarrow +\infty)$ is illustrated in Fig. \ref{fig:thermal}. Since the entropy should always be smaller or equal to the thermal value, we conclude that our approximation must fail near $n\rightarrow 1$ where $S_{A,n}>S_{A,n}^{\rm th}$. The actual entropy is upper bounded by both our result $S_{A,n}$ and by the thermal entropy $S_{A,n}^{\rm th}$, so that it is below both curves in Fig. \ref{fig:thermal}. The crossing of the two curves occur at $n\simeq 1+\frac{c_vT}{S_0}$ at low temperature, which serves as an estimation of where our approximations fail.

\begin{figure}
[t]
\center
\begin{tikzpicture}[baseline={(current bounding box.center)}]
\draw[->,>=stealth] (0pt,0pt)-- (200pt,0pt) node[right]{$t$};
\draw[->,>=stealth] (0pt,0pt)--(0pt,120pt) node[left]{$ \frac{n-1}{n} S_{A,n} $};
\draw[thick, red](0pt,5pt) .. controls(5pt,6pt) .. (10pt,10pt)-- (75pt,75pt) .. controls (80pt,80pt) and (85pt, 85pt).. (90pt,85pt) -- (180pt,85pt);
\filldraw(0pt,85pt) circle (1pt) node[left] {$M \frac{c_v T}{2}$};
\draw[red,<-,>=stealth] (120pt,90pt) -- (140pt,100pt) node[right]{bound};
\draw[thick](0pt,4pt) .. controls(5pt,4pt) .. (10pt,8pt)-- (30pt,28pt);
\draw[dashed, thick](30pt,28pt) .. controls (45pt,40pt) and (60pt, 50pt) .. (90pt, 70pt) .. controls (110pt,83pt) and (120pt,83pt).. (150pt,83pt) -- (180pt, 83pt) node[  right]{slower linear growth};

\draw[dashed, thick, blue](30pt,28pt) .. controls (45pt,35pt) and (70pt, 45pt) .. (90pt, 50pt) .. controls (110pt, 55pt) and (120pt,55pt).. (150pt,55pt) node[right]{sublinear growth};

\draw[<-,>=stealth] (30pt, 25pt)--(60pt,25pt) node[right]{Gaussian correction};
\filldraw (30pt,0pt) circle (1pt) node[below]{$\sim \frac{\alpha_{\mathrm{S}}}{\beta J}$};
\filldraw[red] (80pt,0pt) circle (1pt) node[below]{$\sim M\frac{\alpha_{\mathrm{S}}}{\beta J}$};
\end{tikzpicture}
\caption{A sketch of the possible entropy growth curve in a chain with $M$ sites. The red line is the bound derived using the geometric interpretation with global/collective reparametrizations. It grows linearly until a time scale proportional to $M \frac{\alpha_{\mathrm{S}}}{\beta J}$. The black line denotes the early time weak link result, which contains a linear growth with the same rate, and receives a correction at an $M$-independent value of order$\frac{\alpha_{\mathrm{S}}}{\beta J}$. After that, we have no definitive predictions using the tools we have developed in this paper. It is unclear whether the entropy continue to grow linearly (but presumably with a slower growth rate) until it reaches the late time bound we have derived.   The growth towards this bound may be sublinear, and it is even possible that the entropy saturates at a value sublinear in $M$ as $t\rightarrow \infty$.}
\label{fig: sketch for growing}
\end{figure}
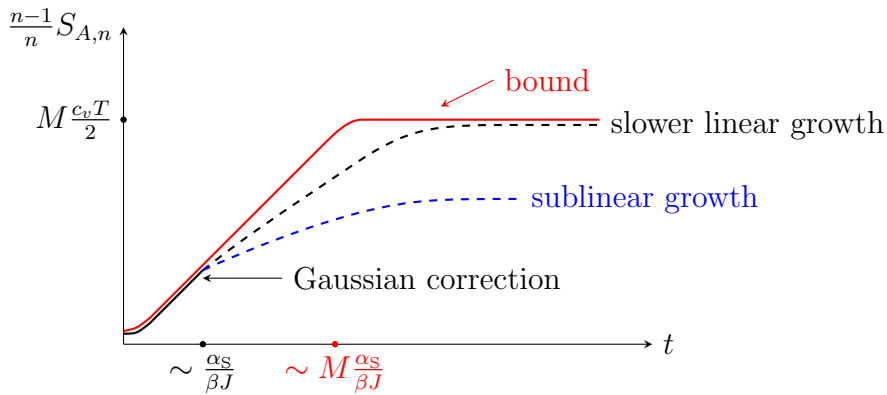

For chains with more than two sites, it is difficult to solve the non-linear equation determining the saddle point of $f_x(\tau)$. However, we can use the same argument above and obtain an upper limit of the R\'enyi entropy. If we consider a uniform ansatz $f_x(\tau)=f(\tau)$, the effective action reduces to the same form as the two-site case, with the parameter $2\alpha_{\mathrm{S}}$ rescaled to $M\alpha_{\mathrm{S}}$ when there are $M$ sites. Therefore the minimization problem can be mapped to the same geometric problem, with the effective coupling parameters $\alpha_{\mathrm{S}},\gamma$ replaced by $M\alpha_{\mathrm{S}}/2,\gamma$. 
In the large $M$ limit, so long as we take the uniform ansatz above, we are in fact closer to the perturbative limit since $\gamma/M\alpha_{\mathrm{S}}\rightarrow 0$ as $M\rightarrow \infty$. The saddle point gives a long time saturation entropy that is simply $M/2$ times the two site value:
\begin{align}
S_{A,n}(\infty)\frac{n-1}{n}=\frac{M c_v T}{4}+\calO(1),~S_{A,n}(0)=\calO(1)
\end{align}
Therefore the entropy grows from area law to volume law, but the final entropy density is still lower than the thermal value. Since this is an upper bound of the actual entropy, we conclude that the chain with generic size $M$ also reaches a pre-thermalized state at long time.  Furthermore, as we noted in Sec. \ref{sec:gaussian}, there is likely a significant correction to the growth rate of entanglement at the time scale $t\sim t_*= \frac{\alpha_{\mathrm{S}}}{\beta J}$, {\it independent from the length of the chain}. This deviation indicates that the upper bound we found for the long chain is not tight. 
There are two possibilities about the fate of entropy growth in a long chain. The final entropy in large $M$ may either be proportional to $M$ or grow slower than $M$. In the latter case, we would consider 
(at least part of the system) 
to be many-body localized. An illustration is given in Fig. \ref{fig: sketch for growing}. Physically we expect that a volume law entropy is more likely, although the growth rate may be quite slow.

\section{Comparison to Holography}\label{sec:holography}
In this section we briefly compare our results to intuition from gauge-gravity duality.    Many features of the SYK model are known to be shared with models of (nearly) $\mathrm{AdS}_2$ gravity:  in particular, a common effective action \cite{stanford1604, jensen, maldacena2016conformal}, and similar thermoelectric transport properties \cite{davison2017thermoelectric}.     Hence it is natural to ask whether the spread of entanglement shares any qualitative features.

Unlike in the SYK model,   in holography it is easier to study von Neumann entropy rather than R\'enyi entropy.   There is a well-known formula for $v_{\mathrm{E}}$ in holography in the TFD quench \cite{hartman2013time,
liu2014entanglement,
liu2014entanglement2, mezei2}.   Applying this formula to geometries with nearly $\mathrm{AdS}_2\times \mathbb{R}^d$ infrared geometries ($d\ge 1$ is required for the interpretation that entanglement is flowing across a spatial surface), we find that \begin{equation}
v_{\mathrm{E}} \propto T.
\end{equation}
Details of this calculation are found in Appendix \ref{app:holography}.   We emphasize that this definition assumes that $v_{\mathrm{E}}$ has been defined as in (\ref{eq:dSEdt}).   This result agrees with our intuition based on dimensional analysis, presented in the introduction, and our early time result from the SYK chain.   However, there are important differences between the late time behavior of the holographic von Neumann entropy $S_{\mathrm{E}}(t)$ and that of the SYK R\'enyi entropy.  For a region of large but finite width $R$ (analogous to the finite length chains studied above), the holographic saturation entanglement is given by \begin{equation}
S_{\mathrm{E}}(\infty) \approx 2s_{\mathrm{th}}R,
\end{equation}
and the saturation time will be approximately \begin{equation}
t_{\mathrm{sat}} \approx \frac{R}{2v_{\mathrm{E}}} \propto \frac{1}{T}.
\end{equation}

This behavior is quite different from the SYK chain.   Strictly speaking, the calculation was performed in a different limit.   In the SYK chain model,  we required that $n-1 \ge 1$ was an integer, and the holographic calculation of the previous paragraph was performed exactly at $n=1$.    While it is difficult to reliably perform a calculation in both models for the same value of $n$, we can make some preliminary comments about the behavior of the holographic calculation for $n>1$.     If we define \begin{equation}
\tilde{S}_{A,n} = n^2 \partial_n \left(\frac{n-1}{n}S_{A,n}\right),
\end{equation}
then \cite{dong} has shown that \begin{equation}
\tilde{S}_{A,n} \propto \mathrm{Area}(\text{brane of tension }\propto n-1),
\end{equation}
where we calculate the length of a cosmic brane of tension $\propto n-1$, stretching between the boundaries of (two-sided) AdS subject to suitable boundary conditions.  Because this brane has finite tension, we must account for its gravitational backreaction.   In $\mathrm{AdS}_2$, the lack of gravitational dynamics means that this backreaction is expected to be very severe.   A preliminary hint for the outcome comes from the following argument:   in nearly $\mathrm{AdS}_2$ geometries, the gravitational dynamics are associated with the movement of the boundary of the near-horizon region \cite{maldacena2016conformal, traversable, kourkoulou2017pure}.  Stretching a brane of finite tension from one side of $\mathrm{AdS}_2$ to the other will thus warp the geometry to bring the two sides closer together.   This is likely analogous to the geometric calculation that we performed in the previous section, although the physical interpretation of the tensionful brane is somewhat different.   In the geometric calculation we did, we observed that for any finite string tension $\gamma$, the backreaction of a tensionful brane stretching between two points on the boundary is so strong that as real time $t\rightarrow \infty$, the length of the brane remains finite:
\begin{align}
\lim_{t\rightarrow \infty}  \cosh D(t) =   \frac{128\alpha_{\mathrm{S}}^2}{\gamma ^2}  + \calO(\gamma^0)
\end{align}
Hence, we expect a qualitative change in the R\'enyi entanglement entropy vs. the von Neumann entanglement entropy, and that the R\'enyi entropy for $n>1$ may saturate at a parametrically smaller value than the von Neumann entropy at $n= 1$.

There are two subtleties to note, which make a precise comparison between holography and the SYK chain difficult.   Firstly, $\tilde{S}_{A,n}$ vanishes when $S_{A,n} \propto n/(n-1)$, as we found in our replica diagonal SYK calculation.    To the extent that SYK can recover the holographic results described above as $n\rightarrow 1$, it is crucial that one finds a replica non-diagonal saddle point of the action (\ref{eq:formalaction}).   Secondly, the calculation of holographic R\'enyi entropy requires calculating the backreaction of tensionful branes in $\mathrm{AdS}_2\times\mathbb{R}^d$, not in purely $\mathrm{AdS}_2$.  Perhaps  additional spatial dimensions modify the intuition about gravitational dynamics in $\mathrm{AdS}_2$ that we presented in the previous paragraph.

\section{Discussion}
\label{sec: conclusion and discussion}

We have initiated a study of the spatial spread of entanglement in the SYK chain model.   Although we were unable to compute the von Neumann entanglement directly, we were able to upper bound the spread of R\'enyi entanglement entropy.   We found that this R\'enyi entropy did not saturate at the expected thermal value, but instead at a parametrically smaller value: 
\begin{equation}
\Delta S_{A,n} \propto c_v T.
\end{equation}
This implies that only the light degrees of freedom in the SYK chain (the reparameterization modes) were able to quickly thermalize.   The bulk of the degrees of freedom, which are responsible for the finite zero temperature  entropy, appear to be `localized' on every site, and unable to quickly thermalize.

Our result casts doubt upon whether the conventional interpretation of $v_{\mathrm{E}}$ as a physical velocity scale at which entanglement propagates is sensible, at least for R\'enyi entropy.    Defined as in (\ref{eq:dSEdt}), we have shown that $v_{\mathrm{E}} \propto T$ for the SYK chain.   $v_{\mathrm{E}} \propto T$ can be justified on dimensional grounds, using the local criticality of the SYK chain:  as length does not scale under renormalization group flows, we expect that $v$ has the dimensions of energy.     However, for R\'enyi entropies, the small saturation value of $\Delta S_{A,n}$ suggests that the speed at which entanglement can spread spatially is actually better thought of as \begin{equation}
\tilde{v}_{\mathrm{E}} = v_{\mathrm{E}} \times\frac{s_{\mathrm{th}}}{ \Delta S_{A,n} } \propto T^0.
\end{equation}
This may seem surprising, as $\tilde{v}_{\mathrm{E}} \gg v_{\mathrm{B}}$, in contrast with the conjecture of \cite{lrbutterfly}.   Of course, there are a few caveats to the direct interpretation of $\tilde{v}_{\mathrm{E}}$ as the correct definition of entanglement velocity.  (\emph{i}) We only know that the early time entropy growth rate $\frac{dS_{A,n}}{dt}\propto T$ in the weakly coupled limit. In the low temperature limit $\gamma >\frac1{\beta J}$, our result is only an upper bound of the true entropy, and so it is possible that the entropy growth rate is much slower, leading to a smaller $\tilde{v}_{\mathrm{E}}$.   (\emph{ii}) It is also possible that indeed $\tilde{v}_{\mathrm{E}}>v_{\mathrm{B}}$, which does not directly violate the inequality $v_{\mathrm{E}}\leq v_{\mathrm{B}}$ since the latter is based on the assumption of thermalization\cite{lrbutterfly, mezei1}.

Thus, it is not straightforward in this model what the ``correct" entanglement velocity is, or even how it scales with temperature.   We contrast this with two dimensional conformal field theories,where the $n=1$ and $n>1$ entropies behave qualitatively similarly in such a quench \cite{hartman2013time}.   In fact, there is already a well-known holographic model where $v_{\mathrm{B}}$ is not the fastest ``infrared" velocity scale.   The theory holographically dual to the planar extremal AdS-Reissner-Nordstr\"om black hole has a speed of sound $v_{\mathrm{sound}} \propto T^0$ \cite{lucasrmp}, even while $v_{\mathrm{B}}\propto \sqrt{T}$ \cite{blake2017diffusion}.   Sound waves are classical hydrodynamic excitations that only exist after the onset of  local thermalization (at the very least in a sector containing the energy-momentum tensor).   We do expect that both $v_{\mathrm{sound}}$ and $\tilde{v}_{\mathrm{E}}$ are smaller than the Lieb-Robinson velocity \cite{liebrobinson}.   While there have been some proposals for bounds and relations between the velocities of scrambling, entanglement, and sound \cite{hartnoll2017bound}, it is clear that they must be made more precise.

Two complementary recent studies of quantum quenches in a single site SYK model have recently appeared \cite{eberlein2017quantum, kourkoulou2017pure},  and both show evidence for rapid thermalization.  In particular, in an extension of the SYK model involving a $\chi^q$ interaction instead of $\chi^4$ interaction in (\ref{eq:SYKHamiltonian}),  \cite{eberlein2017quantum} was able to solve for the non-equilibrium $\langle \chi\chi\rangle$ two-point function exactly at $q=\infty$, in a simple quantum quench involving a change in the Hamiltonian at time $t=0$.   In this limit it was observed that (\emph{i}) this two-point function was instantaneously thermal after the quench,\footnote{Such instantaneous thermalization can occur in the context of holographic Vaidya geometries as well \cite{staessens}, for an instantaneous quench.}  and (\emph{ii}) $\langle\chi\chi\rangle$ appears to come entirely from the light reparameterization modes in the SYK model.     Generalizing our analysis of the two-site SYK chain to a model with finite $q$ reveals that (\emph{i}) the saturation entropy $\mathrm{\Delta} S_{A,n}\propto  q^{-2}$ while (\emph{ii}) $dS_{A,n}/dt$ is approximately $q$-independent.    Hence, the entropy saturation time is $\propto q^{-2}$, which means the light degrees of freedom does thermalize instantaneously in the large $q$ limit.
We expect that correlation functions which are dominated by these light degrees of freedom do not detect the slow thermalization of the `heavy' modes that we have found by studying $S_{A,n}$.

It will also be interesting to compare our results with non-perturbative approaches of computing correlation functions in the Schwarzian action\cite{bagrets2017power,mertens2017solving}. As we discussed in Eq. (\ref{eq: twist twopoint}), the Renyi entropy calculation can be considered as a twist-operator thermal two-point function. In the perturbative limit, the Renyi entropy we obtained in Eq. (\ref{Sfirstorder}) corresponds to a two-point function
\begin{eqnarray}
\left\langle X^\dagger _{A,n}(\tau_1)X_{A,n}(\tau_2)\right\rangle_\beta \propto \left(\sin\frac{\pi}{\beta}\left(\tau_1-\tau_2\right)\right)^{-nN\gamma}
\end{eqnarray}
from which we see that $X_{A,n}$ behaves like a dimension $\frac{nN\gamma}2$ field in the $0+1$-dimensional conformal quantum mechanics. Beyond the perturbative limit, it will be interesting to apply the techniques in Ref.~\cite{bagrets2017power,mertens2017solving} to this heavy operator two-point function problem, as a comparison to our results.

In summary, it appears that the SYK chain  is both maximally chaotic at short times \emph{and} takes a long time to completely thermalize, at least in the special state that we have prepared.   These statements are not inconsistent:  the rapid scrambling of the SYK model comes entirely from the Schwarzian action for the reparameterization modes.  It would be interesting if there are other notable consequences of the remaining, slowly thermalizing, heavy modes.   We have also proposed that a similar prethermalization phenomenon may arise in the maximally chaotic holographic models with $\mathrm{AdS}_2$ infrared geometries.   It would also be worth studying this more closely in the future.

We have left open the possibility that the von Neumann entanglement entropy saturates at the thermal value in our quench setup,  even as we have shown that all higher R\'enyi entropies saturate at a parametrically smaller value.   It would be interesting to resolve this question in future work.   Even were this to occur,  the reduced density matrix of one half of the chain should not appear thermal. Interestingly, it is known that in holographic theories there are examples in which two density matrices have identical von Neumann entropy but distinct Renyi entropies to the leading order of $N$ in the large $N$ limit. For two disjoint regions $A$ and $B$ with their minimal surface also disjointed, the density matrix $\rho_{AB}$ and $\rho_A\otimes \rho_B$ have identical von Neumann entropy to the leading order of $N$ (thus vanishing mutual information $I(A:B)$), but different Renyi entropies\cite{dong}. We do not know whether such a qualitative discrepancy between $n=1$ and $n>1$ R\'enyi entropies signals something more profound about the dynamics of the model.

In conformal field theories in two, three and four spacetime dimensions, $v_{\mathrm{E}}$, $v_{\mathrm{B}}$, $v_{\mathrm{sound}}$ and the speed of light $c$ differ by, at most, about a factor of 2.   The SYK chain is one example of a class of theories where these quantities can be parametrically different.   It will serve as an excellent model for sharpening and making precise a deep yet mysterious relationship between chaos, thermalization, entanglement and hydrodynamics.

\section*{Acknowledgements}

We would like to thank 
Dmitry Bagrets, 
Mike Blake, 
Bowen Chen,
Alexei Kitaev, 
Subir Sachdev,
Douglas Stanford
and
Herman Verlinde 
for helpful discussions.
YG and XLQ are supported by the David and Lucile Packard Foundation. AL is supported by the the Gordon and Betty Moore Foundation's EPiQS Initiative through Grant GBMF4302.  AL also thanks the Aspen Center for Physics, which is supported by National Science Foundation grant PHY-1607611, for hospitality while this work was in progress.

\appendix

\section{Details of the Gaussian approximation}

In this appendix, we present the details of the calculation of the second order correction to $S_{A,n}(t)$ in $\gamma$.

\subsection{Infinite chain}
\label{appendix: infinite chain}

The propagator of the reparametrization field $\epsilon_{x,t}$ is determined by the quadratic action:
\begin{equation}
\frac{\alpha_{\mathrm{S}}}{\beta J} \sum_{|n|\geq 2,p} \frac{1}{2} |n| (|n| + \frac{\beta}{2\pi} 2D(1-\cos p)) (n^2-1) \epsilon_{n,p} \epsilon_{-n,-p}  \label{eq:appendixepsquadratic}
\end{equation}
where we replace $p^2$ by its lattice regularized form $2(1-\cos p)$ as we are going to integrate over the whole Brillouin zone. 
For simplicity, we will neglect the large $N$ factor in the effective action in this appendix, since it is an overall prefactor and only rescales the final answer.  The two-point function of $\epsilon$ that arises from (\ref{eq:appendixepsquadratic}) is
\begin{align}
\langle \epsilon_{x,\tau}\epsilon_{x,0} \rangle& = \frac{\beta J}{\alpha_{\mathrm{S}} } \frac{1}{{M}}  \frac{\beta^2}{(2\pi)^4}  \sum_{p=0,\frac{2\pi}{M}, \ldots, \frac{2\pi}{M}(M-1)}  \sum_{n\geq 2} \frac{2 \cos \frac{2\pi n}{\beta} \tau}{n(n+\frac{\beta}{2\pi} D(1-\cos p) )(n^2-1)} \nn \\
 & \simeq \frac{\beta J}{\alpha_{\mathrm{S}} }  \frac{\beta^2}{(2\pi)^4}   \sum_{n\geq 2}  \int_0^{2\pi} \frac{dp}{2\pi} \frac{2 \cos \frac{2\pi n}{\beta} \tau}{n(n+\frac{\beta}{2\pi} D(1-\cos p) )(n^2-1)} \nn \\
& =  \frac{\beta J}{\alpha_{\mathrm{S}} }  \frac{\beta^2}{(2\pi)^4}  \sum_{n\geq 2}   \frac{2 \cos \frac{2\pi n}{\beta} \tau}{n(n^2-1)} \frac{1}{\sqrt{n \left(n+ \frac{2\beta D}{\pi}\right)}}
\end{align}

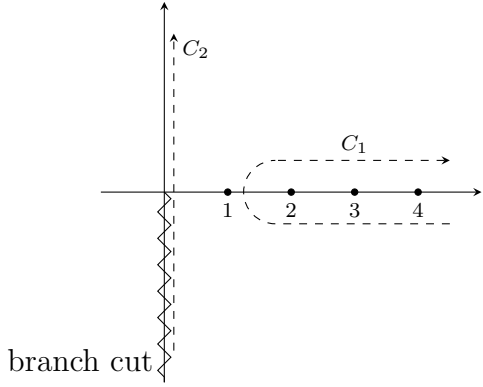
\begin{figure}
[t]
\center
\begin{tikzpicture}[scale=1.2, baseline={(current bounding box.center)}]
\draw[->,>=stealth] (-20pt,0pt) -- (100pt,0pt);
\draw[->,>=stealth] (0pt,-60pt) -- (0pt,60pt);
\filldraw (20pt,0pt) circle (1pt) node[below]{\scriptsize $1$};
\filldraw (40pt,0pt) circle (1pt) node[below]{\scriptsize $2$};
\filldraw (60pt,0pt) circle (1pt) node[below]{\scriptsize $3$};
\filldraw (80pt,0pt) circle (1pt) node[below]{\scriptsize $4$};
\draw[dashed] (90pt,-10pt) -- (35pt,-10pt);
\draw[<-,>=stealth,dashed] (90pt,10pt) -- (35pt,10pt);
\node at(60pt,15pt) {\scriptsize $C_1$};
\node at(10pt,45pt) {\scriptsize $C_2$};
\draw[dashed] (35pt,-10pt) arc (270:90:10pt);
\draw[dashed,->,>=stealth] (3pt,-50pt)--(3pt,50pt);
\draw[decorate, decoration=zigzag] (0pt,0pt)--(0pt,-60pt) node[above left]{branch cut};
\end{tikzpicture}
\caption{Contour deformation $C_1 \rightarrow C_2$. The branch cut (zigzag line) arises from the $x^{-3/2}$ term in the integrand (\ref{eqn: branch cut}) should also be deformed to the left half plane $\Re x\leq 0$, such that the integrand is analytic on the right half $\Re x >0$ except the poles at discrete integer points $x=1,2,3,\ldots$.}
\label{fig: contour deformation for matsubara}
\end{figure}

Now we need to compute the infinite summation over integer $n\geq 2$. Let us denote \begin{equation}
I(\theta) = \sum_{n\geq 2}   \frac{2 \cos n \theta }{n(n^2-1)} \frac{1}{\sqrt{n \left(n+ a\right)}}
\end{equation}
 with $a>0$ and $0\leq \theta < 2\pi$.
 The summation over integers $n$ can be done through the Matsubara trick:
\begin{align}
I(\theta) = \int_{C_1} dx  \frac{1}{e^{-2\pi i x}-1} \frac{e^{-i x \theta} }{(x^2-1)x^{3/2}\sqrt{(x+a)}}  +  \int_{C_1} dx  \frac{1}{1-e^{2\pi i x}} \frac{e^{i x \theta} }{(x^2-1)x^{3/2}\sqrt{(x+a)}} 
\label{eqn: branch cut}
\end{align}
where $C_1$ is an integration contour that winds integer points $2,3,\ldots$ clockwise, see Fig.~\ref{fig: contour deformation for matsubara} for an illustration. The integrand is analytic in the right half plane $\Re x> 0$ except the integer points $x=1,2,\ldots$. Therefore we can deform the contour to $C_2: -i\infty + \epsilon \rightarrow i\infty+ \epsilon $ as shown in Fig.~\ref{fig: contour deformation for matsubara} with the cost of a double pole at $x=1$:
\begin{align}
I(\theta) = &\int_{C_2} dx  \frac{1}{e^{-2\pi i x}-1} \frac{e^{-i x \theta} }{(x^2-1)x^{3/2}\sqrt{(x+a)}}  +  \int_{C_2} dx  \frac{1}{1-e^{2\pi i x}} \frac{e^{i x \theta} }{(x^2-1)x^{3/2}\sqrt{(x+a)}} \nn \\ 
&+ 2\pi i \operatorname{Res} \left(  \frac{1}{e^{-2\pi i x}-1} \frac{e^{-i x \theta} }{(x^2-1)x^{3/2}\sqrt{(x+a)}} + \frac{1}{1-e^{2\pi i x}} \frac{e^{i x \theta} }{(x^2-1)x^{3/2}\sqrt{(x+a)}}   , x=1\right)
\end{align}
The residue can be computed explicitly:
\begin{align}
2\pi i \operatorname{ Res}= 
\frac{(4 a+5) \cos \theta-2 (a+1) (\pi -\theta ) \sin \theta }{2  (a+1)^{3/2}},
\end{align}
while the first two integrals need further treatment.  We notice the integrands diverge near $x=0$ and exponentially decay when going to large imaginary $x$ in the contour $C_2$.  One can show that at large real time $t$, corresponding to large $\mathrm{Im}(\theta)$, it is safe to approximate the integrands by their form near $x=0$:
\begin{align}
I(\theta) &= \int_{C_2} dx  \frac{1}{e^{-2\pi i x}-1} \frac{e^{-i x \theta} }{(x^2-1)x^{3/2}\sqrt{(x+a)}}  +  \int_{C_2} dx  \frac{1}{1-e^{2\pi i x}} \frac{e^{i x \theta} }{(x^2-1)x^{3/2}\sqrt{(x+a)}} \nn \\
&\simeq \frac{1}{\sqrt{a}}  \left( \int_{-\infty}^\infty dx  \frac{ -i }{2\pi x} \frac{e^{ x \theta} }{(ix)^{3/2}}  +  \int_{-\infty}^\infty dx  \frac{-i}{2\pi x} \frac{e^{- x \theta} }{(ix)^{3/2}} \right) \nn \\
&
= \frac{1}{\sqrt{a}}  \left( \int_{0}^\infty dx  \frac{ -i }{2\pi x} \frac{e^{ x (\theta-2\pi)} + e^{-x \theta} }{(ix)^{3/2}}  +  \int_{-\infty}^0 dx  \frac{-i}{2\pi x} \frac{e^{x \theta} + e^{x(2\pi-\theta)} }{(ix)^{3/2}} \right) \nn \\
&= -\frac{1}{\pi \sqrt{2a}} \Gamma\left(-\frac{3}{2} \right) \left( (2\pi-\theta)^{3/2} + \theta^{3/2}   \right) =- \frac{4 }{3\sqrt{2\pi a}} \left( (2\pi-\theta)^{3/2} + \theta^{3/2}   \right) 
\end{align}
Thus, the propagator has an approximate form:
\begin{align}
\frac{(2\pi)^4}{\beta^2}
\frac{\alpha_{\mathrm{S}}}{\beta J} \langle \epsilon_{x,\tau}\epsilon_{x,0} \rangle& \simeq - \frac{4   }{3\sqrt{2\pi a}} \left( (2\pi-\theta)^{3/2} + \theta^{3/2}   \right)  + \frac{(4 a+5) \cos \theta-2 (a+1) (\pi -\theta ) \sin \theta }{2  (a+1)^{3/2}}
\end{align}
where $\theta = \frac{2\pi \tau}{\beta}$ and  $a= \frac{2\beta D}{\pi}$ is small.   We can further simplify the second term to:
\begin{equation}
 \frac{5 \cos \theta -2  (\pi -\theta ) \sin \theta }{2 }.
 \end{equation} 
This step amounts to replacing $\sqrt{x(x+a)}$ by $x$ when $a$ is small.  
We can now evaluate the leading growing term in the Gaussian correction for the effective action at large real time $\tau= \frac{\beta}{2} + i2t $, or $\theta = \pi+ i 4\pi \frac{ t}{\beta}$:
\begin{align}
\left\langle \left( \frac{\epsilon_{x,\theta} -\epsilon_{x,0} }{\tan \frac{\theta}{2} } - \frac{\beta}{2\pi} \left( \epsilon'_{x,\theta} + \epsilon'_{x,0}  \right) \right)^2 \right\rangle 
 \simeq \frac{\beta J}{\alpha_{\mathrm{S}}} \cdot  \frac{\beta^2}{(2\pi)^4}  \cdot \frac{64 \pi}{3 \sqrt{ \frac{2\beta D}{\pi}  }}  \left( \frac{t}{\beta} \right)^{3/2}
\end{align}
Notice the diffusion constant $D$ here can be expressed in terms of the parameter $\gamma= \frac{J_1^2}{8 \pi J^2}$ and $\alpha_{\mathrm{S}}$ using (following \cite{gu2016local}):
\begin{align}
\frac{2\beta D}{\pi} = \beta J \cdot \frac{4 J_1^2}{3\sqrt{2} J^2 \alpha_K} = \beta J \cdot \frac{J_1^2}{8\pi J^2} \cdot \frac{1}{12\alpha_{\mathrm{S}}} =  \frac{\beta J \gamma}{12 \alpha_{\mathrm{S}}}
\end{align}
Therefore, we can express the final minimum of the action in following form:
\begin{align}
 - \frac{1}{2} \left\langle  \left( \frac{\Delta S}{n} \right)^2 \right\rangle \simeq -  \frac{\gamma^2}{8}   \frac{1}{(2\pi)^2}  \frac{\beta J}{\alpha_{\mathrm{S}}} \cdot  \frac{64 \pi}{3 \sqrt{ 
\frac{\beta J \gamma}{12 \alpha_{\mathrm{S}}}
  }}  \left( \frac{t}{\beta} \right)^{3/2} 
  = \frac{2\pi \gamma t}{\beta}  \cdot  \left( - \frac{2}{\sqrt{3} \pi^2 } 
\sqrt{\frac{ J \gamma t}{\alpha_{\mathrm{S}}} }  
   \right)
\end{align}
We factor the $\frac{2\pi \gamma t}{\beta}$ out for easy comparison with the linear $t$ growth term. The formula indicates the linear $t$ growth will receive a correction at time scale:
\begin{equation}
t_* \sim \frac{\alpha_{\mathrm{S}}}{\gamma J}
\end{equation}
as claimed in the main text.

\subsection{Two-sites}
\label{appendix: two sites gaussian}

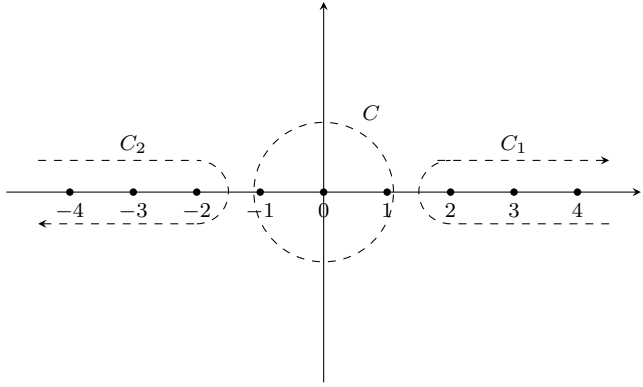
\begin{figure}
[t]
\center
\begin{tikzpicture}[scale=1.2, baseline={(current bounding box.center)}]
\draw[->,>=stealth] (-100pt,0pt) -- (100pt,0pt);
\draw[->,>=stealth] (0pt,-60pt) -- (0pt,60pt);
\filldraw (20pt,0pt) circle (1pt) node[below]{\scriptsize $1$};
\filldraw (40pt,0pt) circle (1pt) node[below]{\scriptsize $2$};
\filldraw (60pt,0pt) circle (1pt) node[below]{\scriptsize $3$};
\filldraw (80pt,0pt) circle (1pt) node[below]{\scriptsize $4$};
\filldraw (0pt,0pt) circle (1pt) node[below] {\scriptsize $0$};
\filldraw (-20pt,0pt) circle (1pt) node[below]{\scriptsize $-1$};
\filldraw (-40pt,0pt) circle (1pt) node[below]{\scriptsize $-2$};
\filldraw (-60pt,0pt) circle (1pt) node[below]{\scriptsize $-3$};
\filldraw (-80pt,0pt) circle (1pt) node[below]{\scriptsize $-4$};
\draw[dashed] (90pt,-10pt) -- (40pt,-10pt);
\draw[<-,>=stealth,dashed] (90pt,10pt) -- (35pt,10pt);
\draw[dashed] (40pt,-10pt) arc (270:90:10pt);
\node at(60pt,15pt) {\scriptsize $C_1$};
\node at(-60pt,15pt) {\scriptsize $C_2$};
\draw[dashed] (-90pt,10pt) -- (-40pt,10pt);
\draw[<-,>=stealth,dashed] (-90pt,-10pt) -- (-40pt,-10pt);
\draw[dashed] (-40pt,-10pt) arc (-90:90:10pt);
\draw[dashed]  (0pt,0pt) circle (22pt) ; 
\node at (15pt,25pt) {\scriptsize $C$};
\end{tikzpicture}
\caption{Contour deformation: the integrand of $I(\theta)$ is analytic in the whole plane except the integer points $x\in \ZZ$. Therefore we can deform the contour $C_1\cup C_2$ that encloses $|x| \geq 2$ integer points to $C$ which only encloses three points $0,\pm 1$.}
\label{fig: contour deformation for two sites}
\end{figure}

The calculations for two sites are much simpler.  Due to the symmetry between the two sites, we can neglect the $p^2$ terms in (\ref{eq:appendixepsquadratic}), and compute the two-point function of $\epsilon$ as
\begin{align}
\langle \epsilon_{x,\tau}\epsilon_{x,0} \rangle& =  \frac{\beta J}{\alpha_{\mathrm{S}} } \cdot \frac{1}{2} \cdot  \frac{\beta^2}{(2\pi)^4}  \sum_{n\geq 2} \frac{2 \cos \frac{2\pi n}{\beta} \tau}{n^2(n^2-1)} 
\end{align}
The extra $\frac{1}{2}$ confirms that the global reparametrization is further suppressed by the system size. Now the evaluation of the summation is much simpler because there is no longer a branch cut of the summand.  
We can complete the contour nicely as shown in Fig.~\ref{fig: contour deformation for two sites}. 
In short, we can compute the infinite sum
\begin{align}
I(\theta): = \sum_{n\geq 2} \frac{2\cos  n \theta }{n^2 (n^2-1)} = \sum_{n\neq 0, \pm 1} \frac{e^{-i n \theta }}{n^2 (n^2-1)} 
\end{align}
by a contour deformation to a simple contour $C$ that only includes three poles (two double poles at $x=\pm 1$ and a triple pole at $x=0$):
\begin{equation}
I(\theta)= \int_{C_1+C_2} \frac{e^{- i x \theta}}{e^{-2\pi i x }-1} \frac{1}{(x^2-1)x^2} = \int_C  \frac{e^{- i x \theta}}{e^{-2\pi i x }-1} \frac{1}{(x^2-1)x^2} 
\end{equation}
Therefore we can easily compute the integral by the residue theorem:
\begin{align}
I(\theta)&=\sum 2 \pi i \left(  \operatorname{Res} (\frac{e^{- i x \theta}}{e^{-2\pi i x }-1} \frac{1}{(x^2-1)x^2}, \{0,\pm 1 \}) \right) \nn\\
&= - \frac{3 \theta ^2-6 \pi  \theta +6 (\pi -\theta ) \sin \theta -15 \cos \theta +2
   \pi ^2-6}{6 }
\end{align}
The correlator has the form \cite{stanford1604}
\begin{align}
\langle \epsilon_\tau \epsilon_0 \rangle = -  \frac{\beta J}{2\alpha_{\mathrm{S}}}  \frac{\beta^2}{(2\pi)^4} \frac{3 \theta ^2-6 \pi  \theta +6 (\pi -\theta ) \sin \theta -15 \cos \theta +2.
   \pi ^2-6}{6 } 
\end{align}
Now we again evaluate the leading growing term in the Gaussian correction by setting $\theta= \pi + i 4\pi \frac{t}{\beta}$
\begin{align}
 - \frac{1}{2} \left\langle  \left( \frac{\Delta S}{n} \right)^2 \right\rangle = - \frac{\gamma^2}{8} \frac{1}{(2\pi)^2} \frac{\beta J}{2\alpha_{\mathrm{S}}} \left( \frac{4\pi t}{\beta} \right)^2 = - \frac{\gamma^2}{4} \frac{\beta J}{\alpha_{\mathrm{S}}} \left( \frac{t}{\beta} \right)^2. \label{eq:lasteqA2}
\end{align}
Again, we see that the linear growth receives a correction at time scale $\frac{\alpha_{\mathrm{S}}}{\gamma J}$.

\subsection{Comparison of the two-site result with the geometric minimization at weak link limit}

\label{appendix: two sites geo}

We can do one further self-consistency check for the two-site problem in the weak link limit.   We now use the geometric interpretation to reproduce the above two-site result.  The strategy is to start with the circle solution, which is a saddle point for the area term, and then expand around the circle solution and find the minimal value when we include the twisted interaction. 

In the geometric picture, we can treat $\alpha_{1,2}$ as function of $y=\cosh D$, and expand the area function around the saddle point:
\begin{align}
\alpha_1^*= 2\pi x, \quad \alpha_2^*= 2\pi (1-x), \quad y^*= 1+ 2 \left( \frac{L}{2\pi} \right)^2 \sin^2 \pi x
\end{align}
The saddle point represents a circle geometrically, which is expected to be the shape with maximal area under constrain. So we must expand the area term to quadratic order in the deviation $y-y_*$:
\begin{equation}
A=A(y^*) -  \frac{1}{2} \cdot \frac{4\pi^6}{L^5} Q(x) (y-y^*)^2.
\end{equation}
The constant term has simple expression $A^*:=A(y^*)=  L + \frac{2\pi^2}{L} - 2\pi$ and
the linear term vanishes since we are expanding around a saddle point for the area. The most important piece is the quadratic term.  We have define a function $Q(x)$ as follows:
\begin{align}
Q(x):=\frac{1}{(\sin \pi x)^4 (1+\pi (1-x) \cot \pi x) (1- \pi x \cot \pi x)} 
\end{align}
and it is clear that $Q(x)$ determines the cost of fluctuations near saddle point $y^*$.  
Note that $Q(x)$ has a minimum at $x=\frac{1}{2}$.
Now the twisted term $\frac{\gamma}{2} \log y$
is a logarithmic function of $y$, which we may also expand to quadratic order:
\begin{eqnarray}
\log y = \log y^* + \frac{1}{y^*} (y-y^*) -\frac{1}{2} \frac{1}{(y^*)^2} (y-y^*)^2.
\end{eqnarray}
Notice that $y^*$ is of order $L^2$; therefore the quadratic term from $\frac{\gamma}{2} \log y $ is of order $\gamma L^{-4}$, while the area term has $L^{-5}$. So expanding around $y$ is useful in the limit $\gamma \ll \frac{1}{L}$, or $\frac{J_1^2}{J^2} \ll \frac{1}{\beta J}$.  In this limit, we get a correction for the saddle point free energy:
\begin{align}
\delta I = \min_{y} \left\lbrace   2\alpha_{\mathrm{S}} \frac{1}{2} \cdot \frac{4\pi^6 }{L^5} Q(x) (y-y^*)^2 + \frac{\gamma}{2} \cdot \frac{1}{y^*} (y-y^*)  \right\rbrace = - \frac{L^5 \gamma^2}{64 \pi ^6 \alpha_{\mathrm{S}}  (y^*)^2 Q(x)}
\end{align}
Notice $y^*= 1+ 2 \left( \frac{L}{2\pi} \right)^2 \sin^2 \pi x$ and we focus on $x$ away from $0$ and $1$, so we can approximate $y^* \sim 2 \left( \frac{L}{2\pi} \right)^2 \sin^2 \pi x $, therefore:
\begin{equation}
\delta I \simeq- \frac{L \gamma^2}{16 \alpha_{\mathrm{S}} \pi ^2} (1+\pi (1-x) \cot \pi x) (1- \pi x \cot \pi x)  \simeq - \frac{L \gamma^2}{4\alpha_{\mathrm{S}}}  \left( \frac{ t}{\beta} \right)^2
\end{equation}
where we have taken $x=\frac{1}{2} - i \frac{2t}{\beta}$ and large real time $t\gg \beta$ to simplify the result. This expression precisely agrees with (\ref{eq:lasteqA2}).

\section{Derivation of the geometric interpretation}
\label{appendix: Geometric interpretation}

In this appendix we provide the derivation of Eq.~(\ref{geo action}).

\subsection{The Schwarzian action term}

The relation between the Schwarzian action and the area enclosed by a closed curve in hyperbolic space has been discussed in \cite{maldacena2016conformal,kitaevIAS}. To make our discussion self-contained, we include a derivation here. We embed our curve in a global Euclidean AdS$_2$ Poincare disk, so that our mapping is a little different from that in \cite{maldacena2016conformal}.

Consider the Poincare disk with metric:
\begin{align}
ds^2= d\rho^2 + \sinh^2 \rho d\theta^2 = \frac{4\left( dr^2 + r^2 d\theta^2 \right)}{(1-r^2)^2} , \quad r= \tanh \frac{\rho}{2} 
\end{align}
Put a curve parametrized by $\tau$: $(r(\tau),\theta(\tau))$, $\tau\in [0,\beta)$ in the hyperbolic disk,  with a large total length $L=\beta J = \frac{\beta}{\epsilon}$, where $J=\frac{1}{\epsilon}\gg 1$. 
The physical time $\tau$ is required to be proportional to the arc length parameter of the curve.  Thus, we fix the induced metric along the curve $g_{\tau \tau}= \frac{1}{\epsilon^2}$, i.e.
\begin{align}
\frac{4(r'^2+r^2 \theta'^2)}{(1-r^2)^2} = \frac{1}{\epsilon^2} .
\label{eqn: constrain for curve}
\end{align}
We always consider the case when the curve is close to the boundary.   Thus, $r\sim 1$ and $r' \ll 1$. The metric constraint then implies that
\begin{align}
r \simeq 1- \epsilon \theta' + \frac{(\epsilon \theta')^2}{2}+ \calO(\epsilon^3)
\end{align}
Using this formula, we can rewrite the area enclosed by the curve as an integral: 
\begin{align}
A=  \int  dr d\theta \frac{4r}{(1-r^2)^2} = \int d\theta\left( \frac{2}{1-r(\theta)^2} -2 \right),
\end{align}
Replacing $\frac{2}{1-r^2}$ using the constraint (\ref{eqn: constrain for curve}), we obtain
\begin{align}
A+4\pi&=\int d\theta \frac{1}{\epsilon} \left( r'^2 +r^2 \theta'^2 \right)^{-1/2} = \int d\theta \frac{1}{\epsilon  r \theta'} \left( 1 - \frac{r'^2}{2r^2 \theta'^2} + \calO(\epsilon^4) \right) \nn\\
&= \frac{1}{\epsilon} \int d\tau \left( 1 + \epsilon \theta' + \frac{\epsilon^2 \theta'^2}{2} -  \frac{\epsilon^2 \theta''^2}{2\theta'^2} \right) + \calO(\epsilon^2) \nn\\
&= L+ 2\pi + \epsilon \int d\tau \left[ \frac{1}{2} \theta'^2 - \frac{1}{2} \left( \frac{\theta''}{\theta'} \right)^2 \right] + \calO(\epsilon^2)
\end{align}
This integral is the same as the Schwarzian action:
\begin{align}
\int d\tau \Sch \left( \tan \frac{\theta}{2} ,\tau \right) 
= \int d\tau \left( \frac{1}{2} \theta'^2 -  \frac{1}{2} \frac{\theta''^2}{\theta'^2} + \left( \frac{\theta''}{\theta'} \right)' \right)
\end{align}
Therefore, we have proven the geometric interpretation of the Schwarzian action \cite{maldacena2016conformal,kitaevIAS}
\begin{align}
\frac{1}{J} \int d\tau \Sch \left( \tan \frac{\theta}{2} ,\tau \right) = A- L + 2\pi
\end{align}
Here $\theta$ is the renormalized reparametrization of time $\theta:= \frac{2\pi}{\beta} f(\tau)$. Each reparametrization $f(\tau)$ determines a angular coordinate $\theta(\tau)$ and further determines the curve  in the hyperbolic disk, by using the constraint (\ref{eq: constrain for curve}).   The Schwarzian action corresponds to the area enclosed by the curve with a fixed length $L=\beta J$.

\subsection{The twist operator term}

Next we need to show the twist operator term $\propto \gamma$ also has a simple interpretation:
\begin{align}
\log \eta_f =\log \cosh D(X_1,X_2),
\end{align}
where $D(X_1,X_2)$ is the distance between two marked points (determined by $\theta_{1,2}=\frac{2\pi}{\beta} f(\tau_{1,2})$) on the curve.  
For this, it is helpful to introduce the embedding coordinates
\begin{align}
\left(X^1, X^2, X^3 \right)= \left( \sinh \rho \cos \theta, \sinh \rho \sin \theta, \cosh \rho  \right) , \quad r= \tanh \frac{\rho}{2}
\end{align}
which live on the hyperboloid $X\cdot X = -1$ in a $3$ dimensional space with metric $(1,1,-1)$.    The distance betwen $X_1$ and $X_2$ is related to the inner product:
\begin{align}
\cosh D(X_1,X_2) &=- \vec{X}_1 \cdot \vec{X}_2 \nn \\
&= \cosh \rho_1 \cosh \rho_2 - \sinh \rho_1  \sinh \rho_2  \left( \cos \theta_1 \cos \theta_2 + \sin \theta_1 \sin \theta_2 \right) \nn \\
& \simeq \frac{e^{\rho_1+\rho_2}}{4} 2 \left( \sin \frac{\theta_1-\theta_2}{2} \right)^2 \simeq 2 \frac{J^2}{ \theta_1' \theta_2'} \left( \sin \frac{\theta_1-\theta_2}{2} \right)^2  \nn \\
&= \frac{J^2 \left[ \sin \frac{\pi}{\beta} \left( f(\tau_1)-f(\tau_2) \right)\right] ^2 }{2 \left( \frac{\pi}{\beta} \right)^2 f'(\tau_1) f'(\tau_2)    }  
= \eta_f \frac{\epsilon_1 \epsilon_2 J^2}{2 }
\end{align}
where we use $\cosh \rho \simeq \sinh \rho \simeq \frac{e^\rho}{2} \simeq \frac{2}{1-r} \simeq \frac{2}{\epsilon \theta'} = \frac{2J}{\theta'}$ in the $1-r\ll 1$ limit. Here $\epsilon_{1,2}$ are UV cut-offs of order $J^{-1}$, therefore $\frac{\epsilon_1 \epsilon_2 J^2}{2}$ is a constant of order $1$ whose accurate determination is unimportant for our purposes. We have assumed that $\frac{\epsilon_1 \epsilon_2 J^2}{2 }=1$ in the main text for simplicity.  Thus we arrive at
\begin{align}
\log \eta_f \simeq \log \cosh D(X_1,X_2).
\end{align}
A different choice of cut-off $\epsilon_{1,2}$ leads to an additional $\calO(1)$ constant, proportional to $\gamma$ and  independent from time $t$.   This does not have any important effect on our results.

\section{Some details of the geometric minimization}
\label{appendix: approximation in geometry}

Here we provide some details for the geometric optimization problem at large real time $t\rightarrow \infty$.   In particular we derive the asymptotic form of the two terms in the action.

\subsection{Solving the constraint}
\label{sec: constrain}
As $t\rightarrow \infty$, our numerics suggests that the angle $\alpha_1 \rightarrow 2\pi$.  We can use this knowledge to derive a constraint between the real and imaginary parts of $\alpha_1$. The starting point is the reality condition \begin{equation}
\Im \left( \frac{\frac{1}{2}-i \frac{2t}{\beta}}{\alpha_1} \sin \frac{\alpha_1}{2} \right) =0.
\end{equation}
This can be rewritten as an equation relating the real and imaginary part of $\alpha_1= 2\pi -\epsilon + i \delta$:
\begin{equation}
\Im \left( \frac{\frac{1}{2}-i \frac{2t}{\beta}}{\alpha_1} \sin \frac{\alpha_1}{2} \right) =0 \Rightarrow 
\frac{4t}{\beta}\left( 
1+ \frac{ \delta \tanh \frac{\delta}{2}  }{ (2\pi -\epsilon) \tan \frac{-\epsilon}{2} }
\right) - \frac{  \tanh \frac{\delta}{2}  }{  \tan \frac{-\epsilon}{2} } + \frac{\delta}{2\pi - \epsilon} =0
\end{equation}
After Taylor expanding the $\tan/\tanh$ functions for small $\epsilon$ and $\delta$ we can solve the equation in leading order:
\begin{align}
\frac{4t}{\beta} \left( 
1- \frac{ \delta^2  }{2 \pi \epsilon  }
\right) + \frac{\delta}{\epsilon}=0
\Rightarrow
\epsilon &= \frac{\delta^2}{2\pi} - \frac{\delta \beta }{4t}
\end{align}

\subsection{Minimization of $I(D)$ at $t \rightarrow \infty$ limit}
\label{appendix: value of c}
\begin{figure}[t]
\includegraphics[width=8cm]{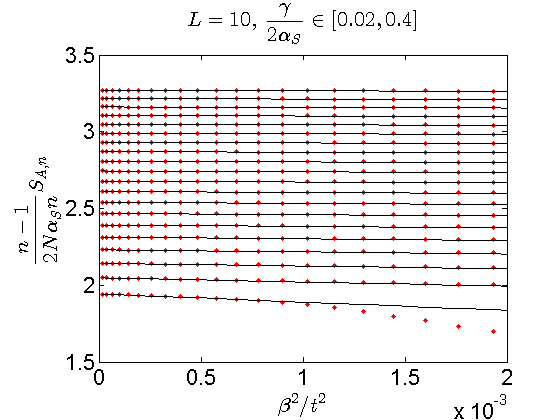}
\centering
\caption{Comparison of the numerical result (red dots) with the long time asymptotic behavior (\ref{longtime limit}) (black lines) for $L=10,~\frac{\gamma}{2\alpha_S}\in[0.02,0.4]$. The curves with higher entropy has higher $\gamma$. \label{fig:longtimefit}}
\end{figure}

In the limit $t \rightarrow \infty$,  we can Taylor expand $A(D)$ and keep the leading order terms:
\begin{equation}
A \simeq  \frac{1}{L} \Re \left( \frac{\alpha_1^2}{\frac{1}{2}- i \frac{2t}{\beta}}  +\frac{4\alpha_1}{\left(\frac{1}{2}- i \frac{2t}{\beta}\right) \tan \frac{\alpha_1}{2}}  \right) + L-2\pi
\end{equation}
Using the constraint $\epsilon = \frac{\delta^2}{2\pi} - \frac{\delta \beta }{4t}$ for $\alpha_1=2\pi - \epsilon +i \delta$, we have:
\begin{align}
A-L+2\pi = \frac{1}{L} \left( 
 \underbrace{ \frac{8\pi \beta }{t\delta}}_{\text{leading}}  
\underbrace{-
\frac{4\pi \beta \delta }{3 t} + \left(\frac{\pi^2}{2}+2\right)\frac{\beta^2}{t^2} - \frac{\pi \beta^3 }{2 \delta t^3}  + \ldots
}_{\text{sub-leading}}
\right)
\end{align}
Since $\delta$ is small, we only need to keep $\frac{16\pi}{\delta t}$ term to minimize the action. 
After $\delta$ is determined variationally, the subleading terms will determine the subleading corrections to the final entropy at late times. We next analyze the $\log$ term:
\begin{align}
\log y &\sim \underbrace{2 \log \left( -\frac{\delta L t}{\sqrt{2} \pi \beta } \right)}_{\text{leading}}   
\underbrace{-\frac{\delta \beta }{4\pi t} + \frac{\beta^2}{8t^2} + \ldots
}_{\text{sub-leading}} 
\end{align}
Now all the terms are explicit function of $\delta$ and we can proceed to find the minimum at leading order:
\begin{equation}
\min_{\delta} \left( - 2\alpha_{\mathrm{S}} \left( \frac{1}{L} \cdot \frac{8\pi \beta }{\delta t} \right) + \gamma \log \delta  \right) \Rightarrow \delta_* \simeq - \frac{16 \alpha_{\mathrm{S}} \pi \beta }{\gamma L t}
\end{equation}
Inserting this value back into the action, we obtain
\begin{align}
I(t) =  \left. \left[- 2 \alpha_{\mathrm{S}} (A-L+2\pi) + \frac{\gamma}{2} \log y\right]\right\vert_{\delta=\delta_*}
\simeq  \gamma \left(1+ \log \frac{8\sqrt{2} \alpha_{\mathrm{S}} }{\gamma} \right) -  c\frac{\beta^2}{t^2}+ \ldots
\end{align}
with $c$ given in Eq. (\ref{eq:cinmain}). We have checked that our numerical results agree well with the analytic result in both the long-time saturation value and the $\frac1{t^2}$ term, as is shown in Fig. \ref{fig:longtimefit}.

\section{Holographic von Neumann Entanglement Velocity}\label{app:holography}
In this appendix we calculate $v_{\mathrm{E}}$ for a holographic model with an $\mathrm{AdS}_2 \times \mathbb{R}^d$ geometry in the IR.   The formula for $v_{\mathrm{E}}$ in a generic planar geometry with metric \begin{equation}
\mathrm{d}s^2 = \frac{L^2}{r^2} \left[\frac{g(r)}{f(r)} \mathrm{d}r^2 - f(r)g(r) \mathrm{d}t^2 + \mathrm{d}\mathbf{x}^2\right]  \label{eq:metricbadcoords}
\end{equation}
is \cite{hartman2013time,
liu2014entanglement,
liu2014entanglement2, mezei2} \begin{equation}
v_{\mathrm{E}} = \left(\frac{r_+}{r_*}\right)^d \sqrt{|(fg)(r_*)}  \label{eq:vEholo}
\end{equation}
where $r_+$ is the location of the (outer) event horizon of (\ref{eq:metricbadcoords}), and  $r_*$ is the solution to \begin{equation}
\frac{2d}{r_*} = \frac{(fg)^\prime(r_*)}{(fg)(r_*)}.  \label{eq:rstarfix}
\end{equation}
which should occur behind the horizon.
Note that $v_{\mathrm{E}}$ is best understood as arising from a calculation of entanglement in a TFD state \cite{hartman2013time}, analogous to the case we studied in the main text.   This calculation may also be done for spatial quenches \cite{liu2014entanglement, liu2014entanglement2},  but in this case it is important that the initial state of the quench has vanishing entropy density; otherwise the formula above is generally modified.

If the matter which sources (\ref{eq:metricbadcoords}) gives rise to an extremal black hole, then at zero temperature $r_+ \rightarrow r_{\mathrm{e}} < \infty$.   (Note that $r_{\mathrm{e}}^{-d} \propto s>0$ \cite{lucasrmp}.)    At a very small but nonzero temperature $T$, we expect that near the horizon, \begin{equation}
f(r)g(r) \approx a(r_{\mathrm{e}}-c_1T-r)^2 - c_2T^2 + \cdots,
\end{equation}
and that $r_+>r_{\mathrm{e}}$ for this new geometry if the specific heat is positive.   The coefficients $c_1$, $c_2$ and $a$ are not independent of $T$ but we will only need the fact that, to leading order in $T$, they are constants.      (\ref{eq:rstarfix}) implies that \begin{equation}
\frac{2d}{r_*} = \frac{-2a(r_{\mathrm{e}}-c_1T-r_*)}{a(r_{\mathrm{e}}-c_1 T - r_*)^2 - c_2 T^2}.
\end{equation}
At small $T$ this equation can only be solved if \begin{equation}
r_{\mathrm{e}} - c_1 T - r_* = bT^2,
\end{equation}
where \begin{equation}
\frac{d}{r_{\mathrm{e}}} = \frac{ab}{c_2}.
\end{equation}
Using (\ref{eq:vEholo}) we conclude that to leading order in $T$ \begin{equation}
v_{\mathrm{E}} \approx \sqrt{c_2T^2 - a(r_{\mathrm{e}} - c_1 T-r_*)^2} = \sqrt{c_2}T + \mathrm{O}(T^2).
\end{equation}
This confirms the scaling that we claimed in the main text.

\bibliography{ref}

\end{document}